\documentclass[aps,pre,reprint,superscriptaddress,longbibliography]{revtex4-1}
\usepackage{amsmath,amssymb,graphicx,bm,epstopdf}

\begin{document}

\title{Nonlinear continuous-wave optical propagation in nematic liquid crystals: interplay between reorientational and thermal effects}

\author{Alessandro Alberucci}
\affiliation{Photonics Laboratory, Tampere University of Technology, FI-33101 Tampere, Finland}
\email{alessandro.alberucci@tut.fi}

\author{Urszula A. Laudyn}
\affiliation{Faculty of Physics, Warsaw University of Technology, PL-00662 Warsaw, Poland}

\author{Armando Piccardi}
\affiliation{NooEL--Nonlinear Optics and OptoElectronics Lab, University "Roma Tre", I-00146 Rome, Italy}

\author{Micha\l{} Kwasny} 
\affiliation{Faculty of Physics, Warsaw University of Technology, PL-00662 Warsaw, Poland}

\author{Bartlomiej Klus} 
\affiliation{Faculty of Physics, Warsaw University of Technology, PL-00662 Warsaw, Poland}

\author{Miros\l{}aw A. Karpierz}
\affiliation{Faculty of Physics, Warsaw University of Technology, PL-00662 Warsaw, Poland}

\author{Gaetano Assanto}
\affiliation{Photonics Laboratory, Tampere University of Technology, FI-33101 Tampere, Finland}
\affiliation{NooEL--Nonlinear Optics and OptoElectronics Lab, University "Roma Tre", I-00146 Rome, Italy}
\email{assanto@uniroma3.it}

%\affil{Optics Lab, Tampere University of Technology, FI-33101 Tampere, Finland}
%\affil{Faculty of Physics, Warsaw University of Technology, PL-00662 Warsaw, Poland}
%\affil{NooEL--Nonlinear Optics and OptoElectronics Lab, University Roma Tre, I-00146 Rome, Italy}

%\affil[*]{Corresponding author: alessandro.alberucci@gmail.com}

%\affil[+]{These authors contributed equally to this work}

\begin{abstract}

Thanks to their unique properties, nematic liquid crystals feature a variety of mechanisms for light-matter interactions. For continuous-wave optical excitations, the two dominant contributions stem from  reorientational and thermal nonlinearities. We thoroughly analyze the competing roles of these two nonlinear responses with reference to self-focusing/defocusing and, eventually, the formation of nonlinear diffraction-free wavepackets, the so-called spatial optical solitons. To this extent we refer to dye-doped nematic liquid crystals in planar cells and  continuous-wave beams at two distinct wavelengths in order to adjust the relative weights of the two responses. The theoretical analysis is complemented by numerical simulations in the highly nonlocal approximation and compared to experimental results. 
\end{abstract}

\date{\today}


\maketitle
\section{Introduction}

Liquid crystals (LCs) are a fascinating state of matter, simultaneously exhibiting  physical properties usually associated with either solids or liquids. This is due, with some exceptions \cite{Kim:2012}, to the decoupling between  positional and  orientational order of the constituting organic molecules: depending on the chemical structure and external physical conditions (temperature, pressure), molecular position and orientation are characterized by a different symmetry \cite{Degennes:1993}. %Nevertheless, due to the fluid-like nature of LCs, coupling between the orientational and positional order can be obtained \cite{Kim:2012}.
In this Paper, we focus on the case in which the position of the molecules is random on the long range, whereas their direction shows a finite degree of orientational order, i.e., on the nematic phase. \\
Composition and arrangement of LCs reflect on their physical and optical properties: LCs usually behave as anisotropic crystals, but with a pointwise direction of the principal axes \cite{Degennes:1993}. At the same time, the high mobility of the LC molecules favors a high degree of tunability through the application of external electric and/or magnetic fields, regardless of their oscillation frequency. Thermotropic LCs are also highly sensitive (including phase transitions) to temperature changes \cite{Tsai:2003}. All these characteristics have been exploited in several areas, the most important being the display technology \cite{Wu:2006}. In optics and photonics, LCs are currently used in many other applications, involving optical signal processing with tunable devices \cite{Maune:2003}, temperature sensing \cite{Moreira:2004}, tunable lasing action \cite{Coles:2010}, spatial light modulation \cite{Zhang:2014}, wavefront manipulation via the Berry phase \cite{Marrucci:2006_1}, among others.   \\
The strong tunability of LCs is present even at optical frequencies. Thus, LCs feature a very strong optical nonlinearity \cite{Tabiryan:1980}. Depending on the excitation, different kinds of nonlinear mechanisms come into play, such as reorientational, thermal, photorefractive, electrostrictive, electronic responses, as well as due to  modulation of the order parameter \cite{Khoo:1995,Wong:1973,Durbin:1981,Peccianti:2010,Piccardi:2011_2}. Hence, on the one hand LCs are an ideal workbench to investigate the interplay between different nonlinearities \cite{Warenghem:2008,Burgess:2009,Laudyn:2015}, on the other hand nonlinear optics helps studying the properties of new LC mixtures, including those doped with dyes \cite{Janossy:1992,Muenster:1997,Ferjani:2006}. \\
In the context of nonlinear optics, materials often exhibit an overall response resulting from the combination of various processes \cite{Marte:1994}. The interplay/competition of nonlinear mechanisms leads to a rich scenario of all-optical phenomena \cite{Maucher:2016}, in many cases adding  extra degrees of tunability and disclosing new ways to optical manipulation and signal processing. Early examples to this extent concerned the study of an intensity-dependent refractive index in Kerr materials with additional higher order contributions \cite{Stegeman:1994,Stegeman:1995,Shim:2012}.
To date, a number of side and higher-order effects have been combined with a self-focusing cubic response as means to stabilize two-dimensional (2+1)D soliton solutions, in time or in space \cite{Kivshar:1995,Quiroga:1997,Mihalache:2006_1,Falcao:2013}, while other examples of the occurrence of two distinct nonlinear responses include spatio-temporal confinement of light into bullets, predicted in nonlocal/noninstantaneous media \cite{Silberberg:1990,Mihalache:2006_2,Burgess:2009}. \\
In this Paper we address the interplay and competition of the two dominant LC nonlinearities when the input is a continuous wave (CW), that is, reorientational and thermo-optic responses to light. We consider a standard planar cell much longer than the diffraction length of the input beam in order to study the mutual role of linear diffraction and nonlinear effects. In fact, both nonlinearities manifest as point-wise changes in the refractive index. At the same time, both of them exhibit a highly nonlocal character, i.e., a light-induced perturbation much wider than the exciting beam \cite{Snyder:1997,Conti:2003,Rotschild:2005}. Since reorientation usually dominates over thermo-optical effects in pure LCs, we consider a specific guest dopant added to the host LC mixture in order to enhance light absorption in a specific  range of wavelengths \cite{Warenghem:2008,Laudyn:2015}. Thereby, employing one wavelength inside and one outside the absorption band of the dye, we can evaluate the interaction of the two responses when simultaneously excited. Using the highly nonlocal approximation \cite{Snyder:1997}, we model the behavior of the two nonlinearities when acting alone or together, considering both the propagation of a single beam  or the interaction of two beams at different wavelengths. Finally, we compare our theoretical findings with experimental measurements.  

%studied both theoretically and experimentally the competition of two kind of nonlinearities in nematic liquid crystal and how their combined action can be used to control the propagation of self-confined beams. 
%With reference to the last issue, due to their physical properties and to their high electrical/optical tunability, liquid crystals are ideal material where to access to different classes of nonlinearities \cite{Khoo:1995} employing relatively simple experimental set-up and low intensities radiation. 

\section{Nonlinear light propagation in Nematic Liquid Crystals}

In the nematic phase (nematic LCs, NLCs), the molecules lack positional order on the long range, but have a high degree of orientational order on macroscopic distances \cite{Degennes:1993}. NLCs are usually featured by a cylindrical symmetry around an axis, termed molecular director $\hat{n}$. The field $\hat{n}$ provides the average direction of the long axes in any given point \cite{Khoo:1995}; optically, $\hat{n}$ is the optic axis of the effective  uniaxial medium. To determine the values of each element of the dielectric tensor, further data on the molecular distribution are required, including the order parameter $S$, which roughly corresponds to the standard deviation of the molecular distribution around $\hat{n}$. The two independent eigenvalues of the dielectric tensor are $n^2_{\bot}$ and $n^2_{\| }$, corresponding to plane waves propagating with phase velocities $c/n_{\bot}$ and $c/n_{\| }$, respectively ($c$ is the speed of light in vacuum) and electric fields oscillating orthogonal or parallel to $\hat{n}$ for $n_\bot$ or $n_\|$, respectively. 
%Standard NLCs are positive uniaxials with optical anisotropy $\epsilon_a=n^2_{\|}-n^2_{\bot} >0$. 
The  thermo-optic effect in NLCs stems from the marked dependence of $S$ on temperature \cite{Degennes:1993}, in turn due to the temperature dependence of the refractive indices $n_{\bot}$ and $n_{\| }$ \cite{Simoni:1997}
\begin{align}
  n_\|(T)   &\approx A- BT + \frac{2(\Delta n)_0}{3} \left(1-\frac{T}{T_{NI}} \right)^\beta, \label{eq:nnor}\\
  n_\bot(T) &\approx A- BT - \frac{(\Delta n)_0}{3}  \left(1-\frac{T}{T_{NI}} \right)^\beta, \label{eq:npar}
\end{align}
where $T_\mathrm{NI}$ is the temperature of the nematic-isotropic transition for a given NLC mixture, whereas $A,\ B, \ \beta$ and $(\Delta n)_0$ are fitting parameters derived from experimental measurements \cite{Li:2004}. From Eqs.~(\ref{eq:nnor}-\ref{eq:npar}) it is clear that $n_\bot$ increases with temperature if $n_\| >n_\bot$ [$(\Delta n)_0>0$ for positive NLCs], whereas $n_\|$ decreases at double rate with respect to $n_\bot$. Fundamental \cite{Derrien:2000} and higher-order solitons \cite{Hutsebaut:2004} based on thermal nonlinearity have been reported in NLCs. \\
For a given direction of the wave vector $\bm{k}$, the eigensolutions of the Maxwell's equations corresponding to $n^2_{\bot}$ and $n^2_{\| }$ are ordinary and extraordinary plane waves, respectively. The ordinary electric field ($o-$wave) is always orthogonal to the director $\hat{n}$ and the phase velocity of the wave is $c/n_o=c/n_{\bot}$. Conversely, the extraordinary ($e-$wave) electric field is coplanar with the wave vector $\bm{k}$ and the director $\hat{n}$, with a phase velocity $c/n_e=c/n_e\left( \theta \right)$ which depends on the orientation angle $\theta$ between $\bm{k}$ and $\hat{n}$ \cite{Simoni:1997} ($n_e \left( 0 \right)=n_\|$):
\begin{equation} \label{eq:ext_index}
  n_e(\theta)= \left(\frac{\cos^2\theta}{n^2_{\bot}}+\frac{\sin^2\theta}{n^2_{\|}} \right)^{-1/2}.
\end{equation}
Moreover, the extraordinary beam propagates in the plane $(\bm{k},\hat{n})$ with a Poynting vector tilted with respect to the wave vector by the walk-off angle $\delta=\arctan[ \epsilon_a \sin2\theta /(\epsilon_a +2n^2_\bot +\epsilon_a \cos 2\theta)]$ \cite{Alberucci:2010_2}, where $\epsilon_a=n^2_{\|}-n^2_{\bot} >0$ is the optical anisotropy, usually positive in NLCs. \\
%\section{Nonlinear optical effects in nematic liquid crystals}
As already stated, the dominant nonlinear optical responses in NLCs excited by CW lasers are  thermal and  reorientational. Thermal nonlinear effects occur, for example, when absorption causes a reduction of the order parameter, with a net decrease (increase) of $n_\|$ ($n_\bot$) \cite{Braun:1993, Warenghem:1998_2} [Eqs.~(\ref{eq:nnor}-\ref{eq:npar})]. Instead, the reorientational nonlinearity originates from collective rotation of molecules induced by light \cite{Khoo:1995,Tabiryan:1980,Durbin:1981}. %Other responses involve fast local changes in the order parameter when using high intensity light pulses \cite{Wong:1973}, as well as ferroelectric or photorefractive effects is suitably doped materials and so on \cite{Khoo:1995}. 
%Hereby we restrict our analysis to nonlocal noninstantaneous nonlinearities excited by (low power) continuous-wave beams, focusing on the two dominant mechanisms taking place in NLCs doped with an absorbing dye: the thermo-optic effect and the molecular reorientation. 
In the case of reorientation, the electric field $\bm{E}$ of a light beam induces a molecular dipole which tends to align to $\bm{E}$. The net result is an electromagnetic torque $\bm{\Gamma} = \epsilon_0 \epsilon_a (\hat{n} \cdot \bm{E}) (\hat{n} \times \bm{E})$. The equilibrium position of the director is determined by the balance between the  torque $\bm{\Gamma}$ and the elastic forces associated with intermolecular links and anchoring conditions, the latter set at the cell edges \cite{Khoo:1995}. When light is purely extraordinary polarized, the all-optical reorientation of the director $\hat{n}$ results in an increase of $\theta$,  leading in turn to an increase in $n_e (\theta)$, as stated by Eq.~\eqref{eq:ext_index}. The net effect is beam  self-focusing. The reorientational nonlinearity is polarization-dependent as well: ordinary input beams orientate the molecules only beyond the Fr\'eedericksz threshold \cite{Durbin:1983}, whereas extraordinary beams can induce nonlinear effects at very low powers \cite{Zeldovich:1980}.\\  
% a small amount of absorbing particles, for example dye compounds or nanoparticles can be added to the NLC to excite different nonlinear phenomena \cite{Janossy:1990,Reznikov:2003,Lucchetti:2006,Piccardi:2010_5,Acreman:2014}.
When an $e-$polarized bell-shaped beam propagates in NLCs, self-focusing at mW excitations can yield light self-confinement and the generation of bright spatial solitons, also termed nematicons \cite{Peccianti:2012}, due to the self-induced transverse refractive index profile. Nematicons are self-trapped beams as well as light induced waveguides for co-polarized signal(s); they can be controlled/routed all-optically, electro-optically, magneto-optically as well as by refractive perturbations, interfaces, boundaries, leading to the implementation of guided-wave circuits for signal addressing and processing \cite{Peccianti:2004,Beeckman:2006,Piccardi:2008,Izdebskaya:2010_2,Assanto:2012,Laudyn:2013,Piccardi:2014,Izdebskaya:2014,Izdebskaya:2017}.
An important feature of self-trapped beams in NLCs is the high degree of nonlocality characterizing the light-induced refractive index potential. As a matter of fact, the light-written index well extends far beyond the beam profile itself, thus preventing the insurgence of catastrophic collapse \cite{Conti:2003,Bang:2002}. In this limit the nonlinear index change $\Delta n$ can be approximated by a parabolic shape $\Delta n=-\phi_2 (x^2+y^2)$, i.e., the system resembles a quantum harmonic oscillator \cite{Snyder:1997}. In writing the expression for $\Delta n$, we assumed $\bm{k}\|\hat{z}$, with the plane $xy$ normal to the wave vector. Nematicons are shape preserving, but in general spatial solitons in NLCs  breathe with excitation dependent periodic oscillations in width and peak intensity \cite{Conti:2004}. The $z$-dependence of the beam width for light propagating in a parabolic index well is given by \cite{Snyder:1997,Alberucci:2015,Alberucci:2016}
\begin{equation} \label{eq:dynamic_breathing}
   \frac{n_0}{2}\frac{d^4 w^2}{dz^4} + 2\phi_2 \frac{d^2 w^2}{dz^2} + 3 \frac{d\phi_2}{dz} \frac{d w^2}{dz} + \frac{d^2 \phi_2}{dz^2} w^2=0,
\end{equation}
with $n_0$ the average refractive index and, in the nonlinear case, $\phi_2$ depending on the light distribution. \\
Here we are interested in the competition between reorientational and thermal effects. In order to study the interplay between these two responses, we employed dye-doped nematic liquid crystals to enhance the optical absorption in a finite region of the spectrum \cite{Warenghem:1998_2}, as detailed in the following section.

\section{Sample geometry and material properties}
\label{sec:geometry}

\begin{figure}[htbp]
\centering
\includegraphics[width=0.47\textwidth]{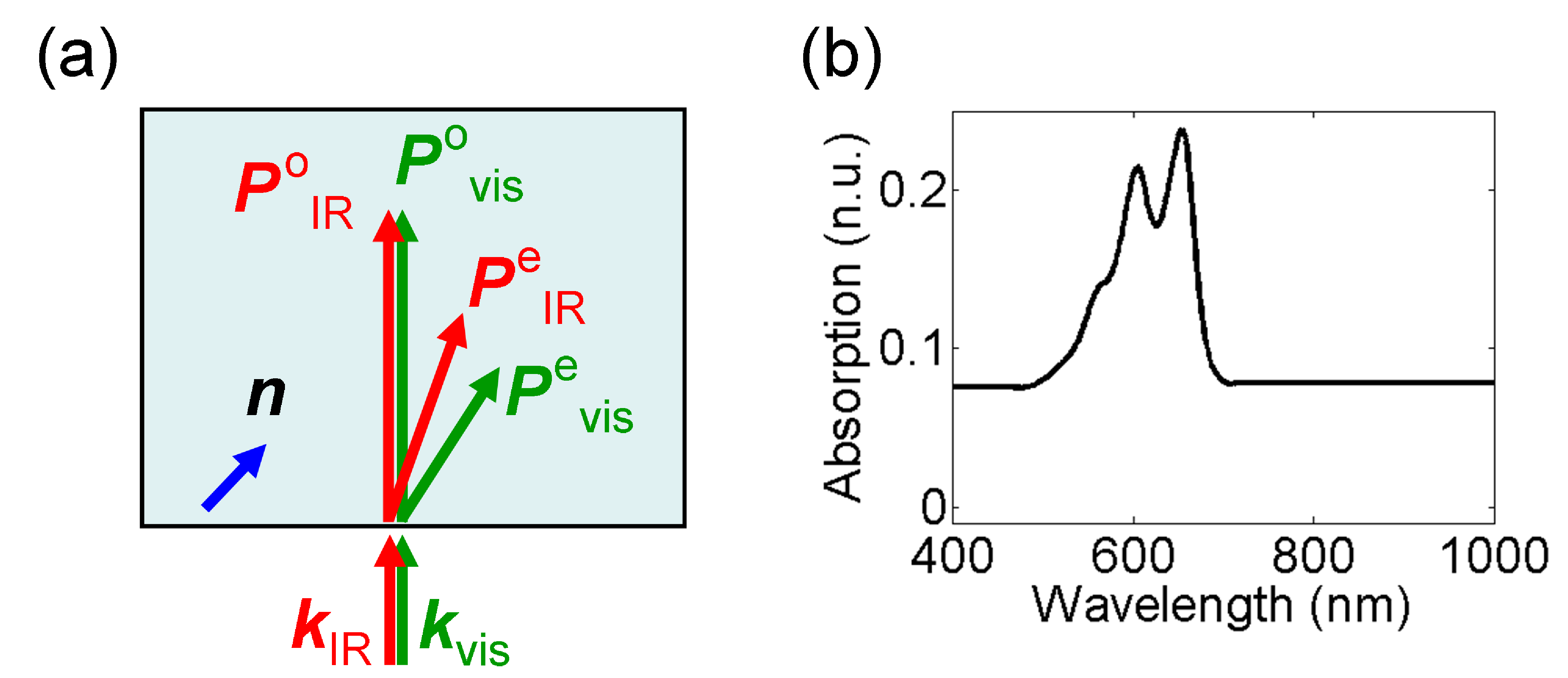}
\caption{(a) Sketch of the NLC sample. Two beams (subscripts \textit{IR} and \textit{vis} for $\lambda =1064$~nm and $\lambda =532$~nm, respectively) are launched collinearly. Inside the anisotropic NLC mixture the Poynting vector of the ordinary components are still parallel while the energy propagates with different walk-off angles. (b) Absorption spectrum of the mixture of 6HBC with 0.1$\%$ of \textit{Sudan Blue} dye.}
\label{fig:cell_sketch}
\end{figure}

The sample  is sketched in Fig.~\ref{fig:cell_sketch}(a). To contain the NLC and maintain the desired molecular alignment, a planar cell was realized with two parallel glass slides separated by $h\approx100~\mu$m, with the inner interfaces mechanically rubbed to yield uniform planar anchoring of the director at $\theta_0=45^\circ$ with respect to the $z$ axis. The cell was filled with the host 6CHBT ($n_{\bot}=1.5021$, $n_{\|}=1.6314$ $@\lambda =1064$~nm and $n_{\bot}=1.52$, $n_{\|}=1.6746$ $@\lambda =532$~nm, room temperature) doped with $0.1\%$ of the \textit{Sudan Blue} dye \cite{Milanchian:2012}, the latter showing a main absorption peak at $\lambda \approx 604~$nm [see Fig.~\ref{fig:cell_sketch}(b)]. \\
Light attenuation in an NLC sample can be described by the (intensity) absorption coefficient  $\alpha_\mathrm{ov}$:
\begin{equation}
\alpha_\mathrm{ov}=\alpha_\mathrm{el} + \alpha. \label{eq:definition_alpha}
\end{equation}
The term $\alpha_\mathrm{el}$ corresponds to the Rayleigh-like scattering, thus implying no changes in  temperature. Conversely, the term $\alpha$ accounts for inelastic scattering, responsible for warming up the sample. When a resonant dye is added to the NLC host, in the limit of small concentrations only the term $\alpha$ changes, allowing to control the amount of heat generated by light in the NLC. In addition, the strong NLC anisotropy yields -in general- a dichroic response, with $\alpha_\mathrm{ov}$ depending on the polarization. \\
We used two CW laser beams of different wavelengths to change the relative weight of the two nonlinear responses. The beam at $\lambda =1064$~nm, away from the dye resonance, excites reorientation. The second beam, with wavelength $\lambda=532$~nm is within the absorption band of \textit{Sudan Blue} to enhance the thermal nonlinearity \cite{Warenghem:2008}. The two beams are coupled into the sample by focusing them with a microscope objective to a waist of $\approx 3~\mu$m at the input section $z=0$. The beam evolution is observed by collecting the out-of-plane scattered light with a CCD camera and a microscope.

\section{Comparison between reorientational and thermal nonlinearities}

\subsection{Reorientational nonlinearity}

The most known all-optical effect in NLC for CW excitations is the reorientational nonlinearity. %Let us define the angle $\theta$ between the wave vector $\bm{k}$ and the director $\hat{n}$, supposing $\bm{k}$ not parallel to $\hat{z}$. 
Describing the  director distribution by the angle $\theta$, molecular reorientation in the single elastic constant approximation obeys \cite{Degennes:1993}

\begin{equation}   \label{eq:reor}
  \nabla^2\theta + \frac{\epsilon_0 \epsilon_a(T)}{4K(T)}\sin[2(\theta-\delta(\theta,T))]|E|^2=0,
\end{equation}

where $E$ is the slowly varying envelope of the propagating field.  %and $\delta=\arctan\{ \epsilon_a \sin(2\theta) /\left[\epsilon_a +2n^2_\bot +\epsilon_a \cos (2\theta)\right]\}$ is the optical walk-off.
In Eq.~\eqref{eq:reor} we wrote explicitly the dependence of the material parameter ($\epsilon_a$ and $K$) on the temperature $T$. Eq. \eqref{eq:reor} is valid for a linear input polarization, with the field parallel to $\hat{x}$ (the ordinary component in the unperturbed NLC, with reorientation in the $xz$ plane) or to $\hat{y}$ (extraordinary component, reorientation in $yz$). As well known, the ordinary wave is subject to the optical Fr\'eedericksz threshold, that is, the director starts to rotate only beyond a given optical power \cite{Durbin:1981,Braun:1993}; conversely, the extraordinary component undergoes  threshold-less reorientation, allowing for the observation of self-focusing even at modest powers. In particular, for small reorientations ($\theta=\theta_0+\psi$ and $\psi\ll\theta_0$) we can assume $\theta\approx \theta_0$ in the \textit{sine} term in Eq.\eqref{eq:reor}, the latter thus becoming a Poisson's equation, linear in beam intensity $I\propto|E|^2$. Hereafter we will proceed in the small reorientation approximation when solving Eq.~\eqref{eq:reor}, consistently with the configuration shown in Section \ref{sec:geometry} and described by $\theta_0=\pi/4$ \cite{Alberucci:2010_2}. 

\subsection{Thermal nonlinearity}

Simultaneously to the torque exerted on the NLC molecules, the beam heats the sample through absorption. Neglecting convection \cite{Beeckman:2005}, the temperature fulfills a Poisson's equation 
\begin{equation} \label{eq:T}
  \nabla^2 T = -\frac{\alpha^j n^j}{2\kappa Z_0}|E|^2 \ \ (j=\text{ext},\text{ord}),
\end{equation}
where $Z_0$ is the vacuum impedance. In Eq.~\eqref{eq:T} both the optical absorption $\alpha^j$ and the refractive index $n^j$ change according to the polarization (with ordinary index $n_\bot$ or extraordinary index given by Eq.~\eqref{eq:ext_index}) of the propagating wave. In writing Eq.~\eqref{eq:T} we neglected the spatial anisotropy of the thermal conductivity $\kappa$ in NLC.

\subsection{Interplay between heat flow and molecular reorientation}

Equations~(\ref{eq:reor}-\ref{eq:T}) allow us to compute the nonlinear perturbation of the temperature distribution and the director field. Once $T$ and $\theta$ are known, Eqs.~(\ref{eq:nnor}-\ref{eq:ext_index}) allow to describe the overall index well -$n_e(\theta,T)$ or $n_o(T)$ according to the input polarization- induced by light. \\
The relative weight of the two nonlinearities and their consequent interplay can be addressed assuming a Gaussian intensity profile $I\propto \exp[-2(x^2+y^2)/w^2]$ and neglecting the derivative in the propagation direction (i.e., the Poynting vector) in Eqs.~(\ref{eq:reor}-\ref{eq:T}): in this limit a closed form solution can be found for the Poisson's equation. The maximum transverse reorientation  $\theta_m$ and temperature $T_m$ (on beam axis) are then \cite{Alberucci:2007} 
\begin{align}
  \theta_m(\theta_0,w,P,T_m) &= \theta_0 + \nonumber \\
	 &{C_\theta(\theta_0,T_m) P } \sum_{l=0}^\infty{\frac{1}{2l+1} \text{erfc}\left[\frac{\pi(2l+1)w}{2\sqrt{2}L_x} \right]} , \label{eq:thetam} \\
	T_m(T_0,w,P) &= T_0 + C_T^j P \sum_{l=0}^\infty{\frac{1}{2l+1} \text{erfc}\left[\frac{\pi(2l+1)w}{2\sqrt{2}L_x} \right]}, \label{eq:Tm}
\end{align}
where $C_\theta(\theta_0,T_m)=\frac{\epsilon_0  \epsilon_a(T_m) Z_0  \sin[2(\theta_0-\delta_0(T_m))]}{8 n_e(\theta_0,T_m)\cos^2\delta_0(T_m) K(T_m)}$, $C_T^j=\frac{\alpha^j}{4\kappa}$, $T_0$ is the room temperature and $P$ is the beam power. Noteworthy, neither $\theta_m$ nor $T_m$ depend explicitly on the wavelength. 
We stress that, if the boundary conditions applied to Eqs.~(\ref{eq:reor}-\ref{eq:T}) are the same \footnote{In actual cases this is valid, at least to first approximation: the temperature is assumed equal to $T_0$ at the edges of the sample, whereas the optical reorientation $\psi$ vanishes at the interfaces under strong anchoring conditions.}, the spatial profile of the nonlinear perturbation is the same for both the nonlinear mechanisms. This is true whenever the anisotropy in the elastic properties and in the thermal conductivity of the NLC is neglected. In other words, the ratio between reorientational and thermal contributions does not vary point by point, at least for small variations of $T_m$ and $\theta_m$ across the sample. A key parameter is the magnitude of the parabolic index well $\phi_2$, unambiguously determining the intensity distribution of a single beam in the highly nonlocal limit. A simple analytical relationship between $\phi_2$ and the beam intensity peak can be easily found by a Taylor's expansion \cite{Conti:2004}. For the reorientational and the thermal cases we respectively find
\begin{align} \label{eq:phi2_reor}
   \phi_{2,\theta}(\theta_0,w,P,T_m) &=\frac{1}{\eta} \left.\frac{dn_e}{d\theta}\right|_{T_m,\theta_0} \frac{2C_\theta(\theta_0,T_m)}{\pi} \frac{P}{w^2},  \\
	 \phi_{2,T}^j(\theta_0,w,P,T_m) &= \frac{1}{\eta} \left.\frac{dn^j}{dT}\right|_{T_m,\theta_0} \frac{2C_T^j}{\pi} \frac{P}{w^2}  \ (j=\text{ext},\text{ord}),  \label{eq:phi2_thermal}
\end{align}
with $\eta$ a fit coefficient (equal to 2) introduced to improve the matching between model and exact solutions \cite{Ouyang:2006,Alberucci:2014}.\\
The two quantities $\phi_{2,\theta}$ and $\phi_{2,T}$ can be substituted into Eq.~\eqref{eq:dynamic_breathing} to find how the beam radius varies along $z$. Assuming $\phi_2=\Omega/w^2$ in agreement with Eqs.~(\ref{eq:phi2_reor}-\ref{eq:phi2_thermal}), the beam width evolves according to \cite{Alberucci:2016}
\begin{equation}
  \frac{d^4 w^2}{dz^4} + \frac{2\Omega}{n_0 w^2}\frac{d^2 w^2}{dz^2} - \frac{2\Omega}{n_0 w^4} \left(\frac{dw^2}{dz}\right)^2=0.
\end{equation}
Let us now discuss the character -focusing or defocusing- of the two nonlinearities with respect to the initial temperature $T_0$ of the NLC layer. The coefficient $\phi_{2,\theta}$ is always positive because reorientation increases the refractive index. Conversely, $\phi_{2,T}^j$ can be either positive or negative, in agreement with Eqs. (\ref{eq:nnor}-\ref{eq:npar}). In particular, $\phi_{2,T}^{\text{ord}}$ is negative, whereas $\phi_{2,T}^{\text{ext}}$ changes its sign as the director orientation $\theta$ varies via the coefficient $dn^{\text{ext}}/dT=dn_e/dT$, with $\theta$ changing the relative weight of $n_\bot$ and $n_\|$ in determining $n_e(\theta)$ [see Eq.~\eqref{eq:ext_index}] \cite{Simoni:1997}. Since for small anisotropy $\epsilon_a$ we can write
\begin{equation}  \label{eq:ne_approx}
 n_e(\theta)\approx n_\bot + \frac{\epsilon_a}{2n_\bot}\sin^2\theta, 
\end{equation}
Equations~(\ref{eq:nnor}-\ref{eq:npar}) yield
\begin{equation}  \label{eq:sign_thermal}
  \frac{dn_e}{dT}= -B - \left[ \frac{(\Delta n)_0 \beta}{3T_{NI}}\left(1-\frac{T}{T_{NI}} \right)^{\beta-1}\right] \left(2 -3 \cos^2\theta_0 \right).
\end{equation}
Equation (\ref{eq:sign_thermal}) states that, in our configuration with $\theta_0=\pi/4$,  the thermo-optic response  for the extraordinary component is defocusing when $B>0$. When $B<0$ the response is defocusing if $T>\left[1-\left(\frac{6T_\mathrm{NI}|B|}{\beta (\Delta n)_0} \right)^{1/(\beta-1)} \right]T_\mathrm{NI}$ is satisfied. For example, at $\lambda=1064$~nm in the mixture 6CHBT with $\theta_0=\pi/4$, Eq.~\eqref{eq:sign_thermal} is negative only for temperatures above 307K. Finally, the magnitude of the nonlinear effects increases as the temperature approaches the transition temperature $T_\mathrm{NI}$ \cite{Simoni:1997}. 

\subsection{Ordinary polarization}
\begin{figure}
\includegraphics[width=0.47\textwidth]{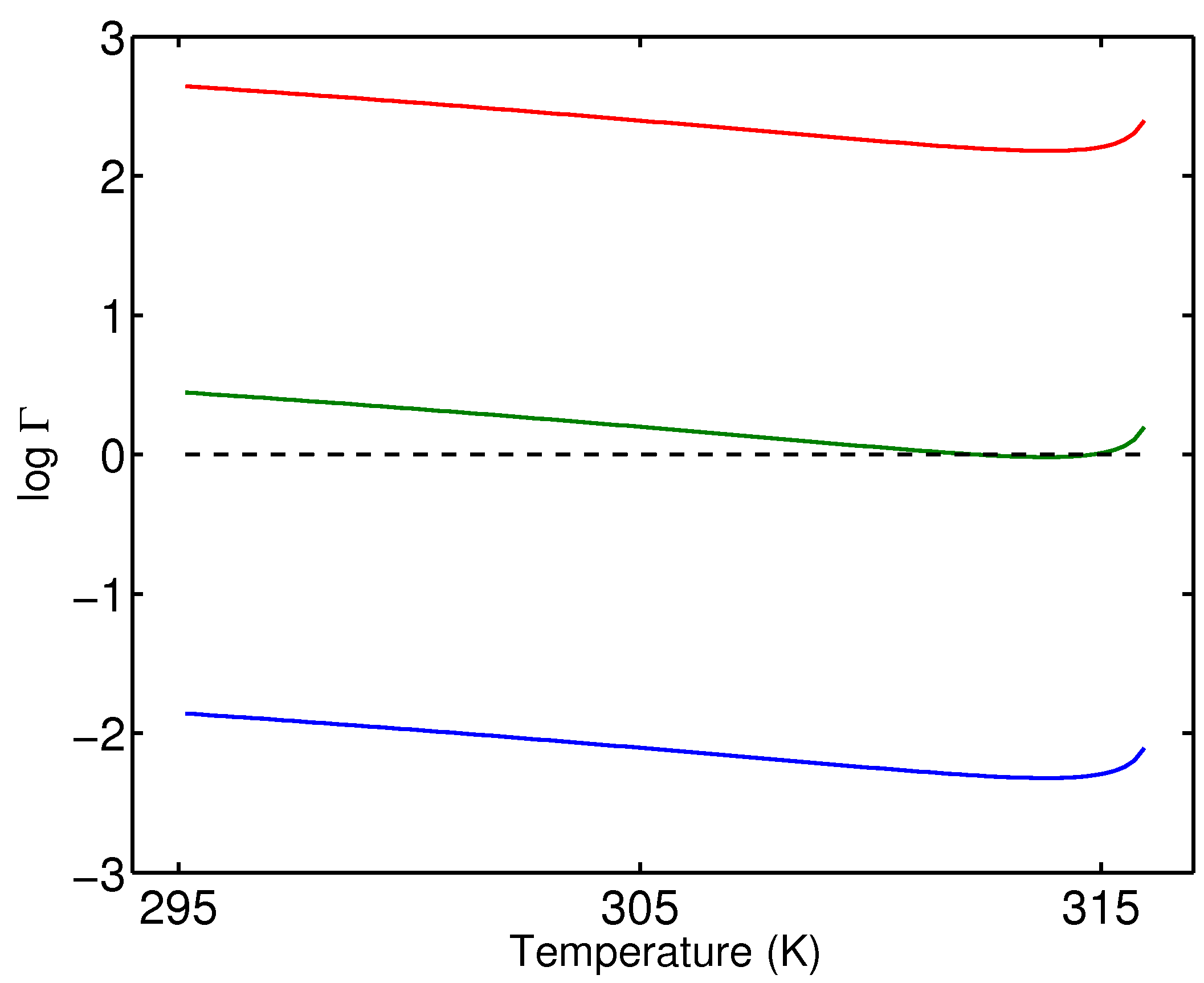}
\caption{Plot of $\log(\Gamma)$ versus temperature for $\alpha^\text{ord}=10^{2},\ 10^{3}$ and $9\times 10^{3}~$m$^{-1}$, from lower to upper curves, respectively. Here $\lambda=532~$nm, $\kappa=1.6\times 10^{-1}$~Wm$^{-1}$K$^{-1}$ and $T_0=22^\circ$C.}
\label{fig:ratio_threshold}
\end{figure}

The ordinary case is quite simple: below the Fr\'eedericksz threshold, only the thermal nonlinearity is active, thus light  undergoes self-focusing (defocusing) when $dn_\bot/dT>0$ ($dn_\bot/dT<0$) \cite{Piccardi:2011_2}, with  beam dynamics  determined by Eqs.~(\ref{eq:Tm}) and (\ref{eq:phi2_thermal}).  Reorientation occurs above the threshold, so  both the nonlinear mechanisms take place at the same time. A crucial issue is whether reorientation or nematic-isotropic transition takes place first. % (close to the nematic-isotropic transition it is always $dn_\bot/dT>0$). 
To answer, let us assume a beam invariant along $z$; the power $ P_\mathrm{th}$ corresponding to the Fr\'eedericksz threshold is found from \eqref{eq:thetam} by setting $\theta_0=0$.  The optical threshold power $ P_\mathrm{th}$ is then \cite{Alberucci:2014_1}
\begin{align}
   &P_\mathrm{th}(T_m)= \nonumber \\
	&\frac{\pi n_\bot(T_m) K(T_m)}{\epsilon_0 \epsilon_a(T_m) Z_0 } \left[\sum_{l=0}^\infty  \frac{1}{ 2l+1} \text{erfc}\left( \frac{\pi (2l+1) w}{2\sqrt{2}L_x} \right)\right]^{-1}. \label{eq:Pth}
\end{align} 
Similarly, starting from \eqref{eq:Tm}, the power $P_\mathrm{NI}$ required to reach the nematic-to-isotropic transition is 
\begin{equation} 
  P_\mathrm{NI}=\frac{4 \kappa (T_\mathrm{NI}-T_0)}{ \alpha } \left[\sum_{l=0}^\infty  \frac{1}{ 2l+1} \text{erfc}\left( \frac{\pi (2l+1) w}{2\sqrt{2}L_x} \right)\right]^{-1}. \label{eq:PNI}  
\end{equation}
As expected, the ratio $\Gamma=P_\mathrm{th}/P_\mathrm{NI}$, determining which phenomenon occurs first, does not depend on the beam width $w$ but only on material parameters, the latter depending on wavelength and temperature. The logarithm of the ratio $P_\mathrm{th}/P_\mathrm{NI}$ is graphed in Fig.~\ref{fig:ratio_threshold} versus  sample temperature for three values of absorption. When $\Gamma$ is lower than 1, the threshold is overcome before the transition to the isotropic phase. Conversely, when $\Gamma>1$ the Fr\'eedericksz transition is preceded (hence, washed out) by the phase transition from the nematic to the isotropic state. 

\subsection{Extraordinary polarization}
The extraordinary polarization case is more complicated tham the previous one due to direct competition of two nonlinearities with opposite signs [see Eq.~\eqref{eq:sign_thermal}]. The dominant term can be found by a direct comparison between Eq.~\eqref{eq:phi2_reor} and Eq.~\eqref{eq:phi2_thermal}. The overall nonlinear effect will be focusing when
\begin{equation} \label{eq:comparison_phi2}
   {C_\theta(\theta_0,T_m)}  \left.\frac{dn_e}{d\theta}\right|_{T_m,\theta_0} >
	    -{C_T} \left.\frac{dn_e}{dT}\right|_{T_m,\theta_0} .
\end{equation} 
Using Eqs.~(\ref{eq:ne_approx}) and (\ref{eq:sign_thermal}), conserving only the lowest order terms in $\epsilon_a$, Eq. (\ref{eq:comparison_phi2}) can be recast as 
\begin{align}  \label{eq:dominant_NL}
   & \frac{ \epsilon^2_a(T_m) \sin^2(2\theta_0)}{ 16 c n^2_\bot(T_m) K(T_m) } + \nonumber \\ & \frac{\alpha^{\text{ext}}\left(2-3\cos^2\theta_0 \right)} {4\kappa} \frac{(\Delta n)_0 \beta}{3T_{NI}}\left(1-\frac{T_m}{T_{NI}} \right)^{\beta-1}  >   \frac{\alpha^{\text{ext}}B} {4\kappa}.
\end{align}

\begin{figure}
\includegraphics[width=0.47\textwidth]{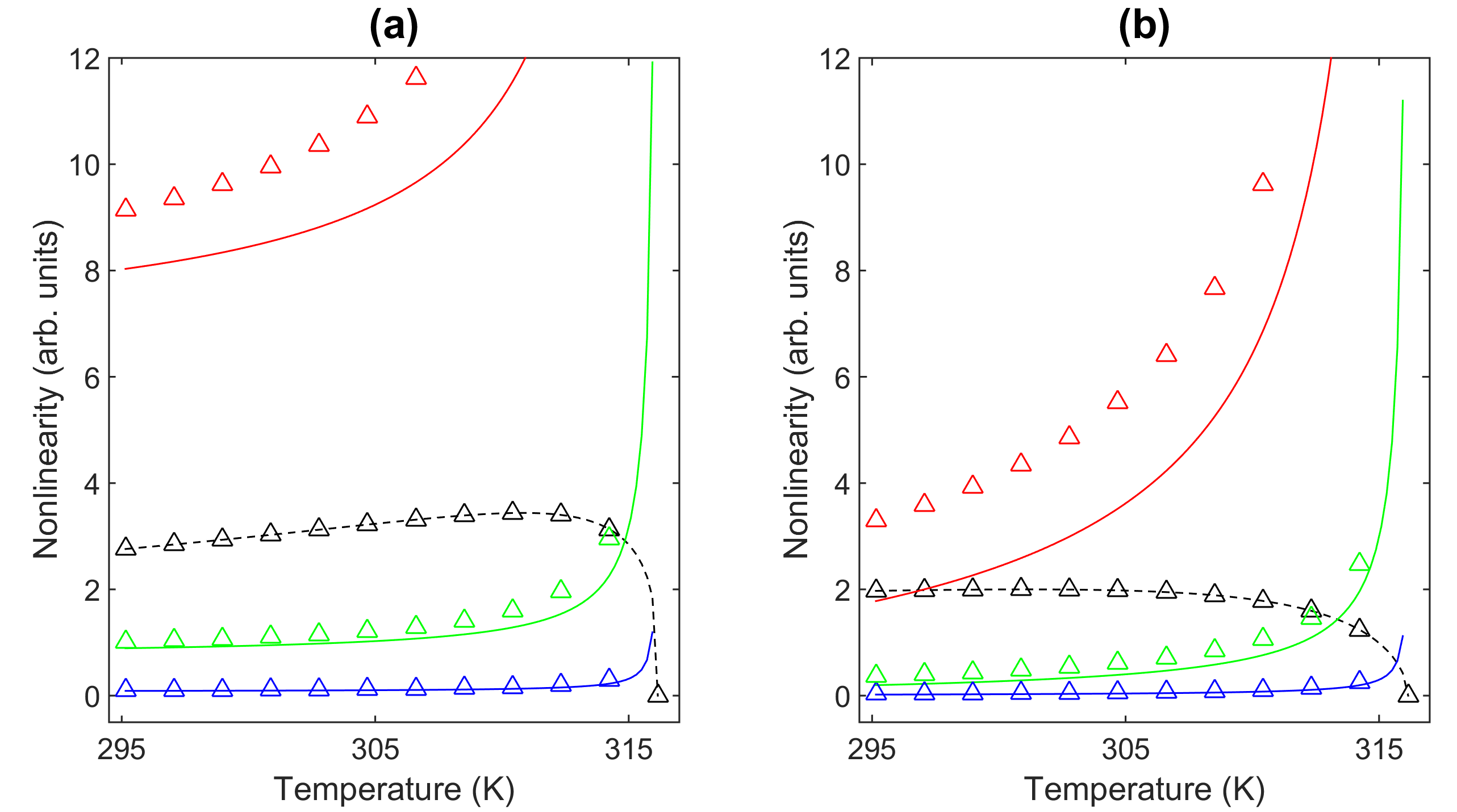}
\caption{ Competition between reorientational and thermal nonlinearities for extraordinary waves in 6CHBT when (a) $\lambda=$532~nm and (b) $\lambda=1064~$nm, evaluated from Eq.~\eqref{eq:dominant_NL} with $\theta_0=\pi/4$. Dashed black line and solid lines are the left-hand side and the right-hand side of Eq.~\eqref{eq:dominant_NL} versus sample temperature, respectively. The thermal nonlinearity is computed for $\alpha^\text{ext}=1\times10^{2}$~m$^{-1}$ (bottom blue solid lines), $1\times10^{3}$~m$^{-1}$ (middle green solid lines) and $9\times10^{3}$~m$^{-1}$ (top red solid lines). Triangles correspond to the exact expressions, Eqs.~(\ref{eq:phi2_reor}-\ref{eq:phi2_thermal}). }
\label{fig:comparison_NL_ext}
\end{figure}

%\left[B + \frac{(\Delta n)_0 \beta}{3T_{NI}}\left(1-\frac{T_m}{T_{NI}} \right)^{\beta-1} \right]  }
Typical results are illustrated in Fig.~\ref{fig:comparison_NL_ext} for parameters corresponding to the mixture 6CHBT. The overall nonlinearity will be defocusing (i.e. thermal heating prevailing on molecular reorientation) in the presence of a large absorption. When the absorption reduces, reorientation becomes dominant. The relative weight of the two  mechanisms depends on  temperature: thermal effects undergo a steep increase close to $T_\mathrm{NI}$, thus  reorientation dominates over heating for temperatures far below the transition. The temperature interval  where the torque prevails on heating gets wider as absorption diminishes, becomes very narrow for undoped NLC ($\alpha^\text{ext}_\mathrm{ov}\approx\alpha^\text{ext}_\mathrm{el}\approx  10^{2}$ m$^{-1}$). Figure~\ref{fig:comparison_NL_ext} also shows that reorientational effects are lower in the green than in the IR, if the thermal absorption is supposed to be the same at the two wavelengths.  

\subsection{Scalar fundamental soliton in dye-doped 6CHBT}

\begin{figure}
\includegraphics[width=0.47\textwidth]{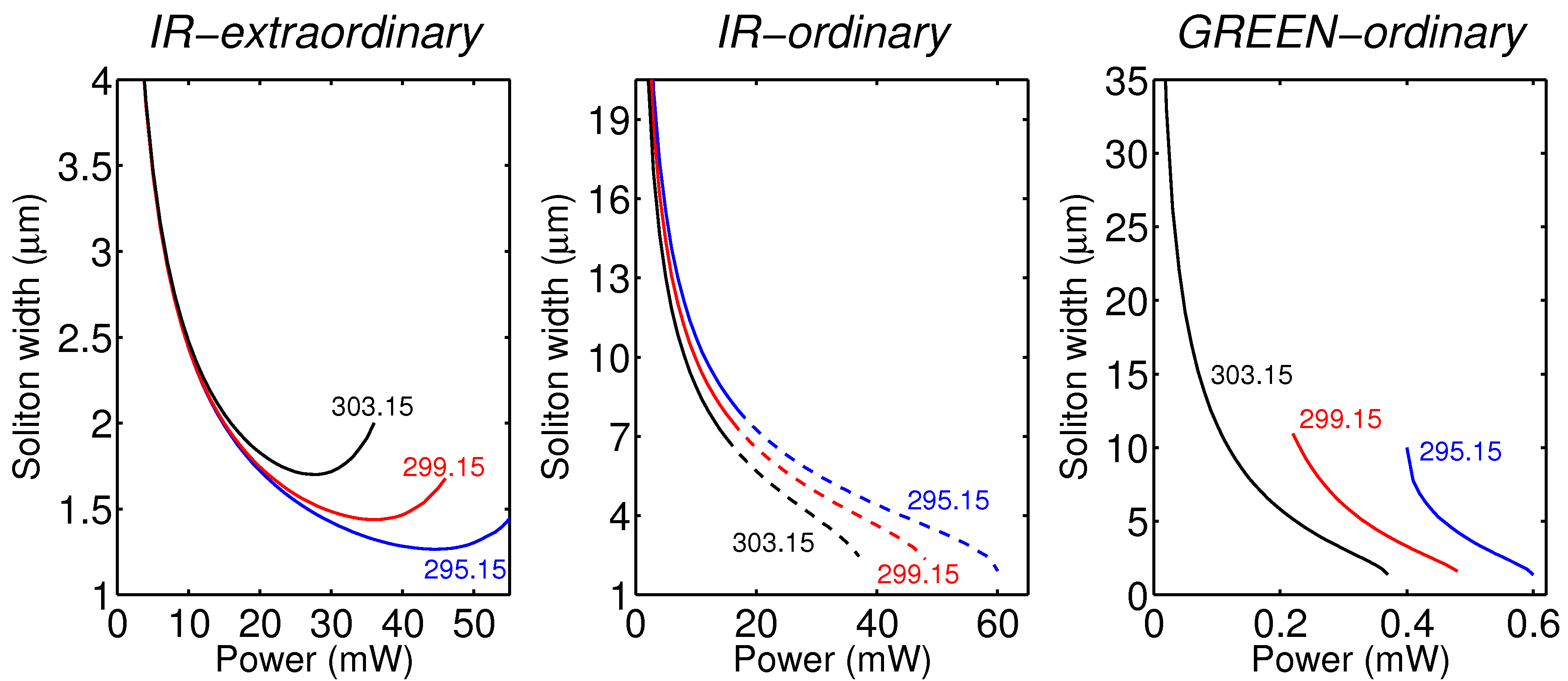}
\caption{Soliton width versus input power for different input polarizations and wavelengths in doped 6CHBT with $\theta_0=\pi/4$ and absorption $\alpha$ of $10^{2}$~m$^{-1}$ (IR) and of 9$\ \times\ 10^{3}$~m$^{-1}$ (green), respectively. The marked values are the initial temperatures $T_0$ in Kelvin. The dashed lines in the center panel correspond to overcoming the Fr\'eedericksz threshold, inhibiting the observation of a purely thermal soliton.}
\label{fig:soliton_w_vs_P}
\end{figure}

Let us now specialize our general considerations to the sample we used in experiments and detailed in Section \ref{sec:geometry}. For the sake of simplicity, we neglect NLC dichroism and take $\alpha$ independent from the wave polarization. We can estimate a strong absorption  $\alpha\gg \alpha_\mathrm{el}$, equal to about 9$\times 10^{3}$~m$^{-1}$, for $\lambda=532~$nm due to the presence of the dopant. At $\lambda=1064$~nm  no resonance with the dye is excited (negligible $\alpha$) and  $\alpha_\mathrm{ov}\approx 10^{2}$~m$^{-1}$, as in standard NLC. Figure~\ref{fig:soliton_w_vs_P} graphs the width of the fundamental (single-humped) soliton versus input power of the beams at the two wavelengths. These solitary waves are scalar, i.e., they encompass only one polarization (ordinary or extraordinary) at a given wavelength. The soliton features have to account for both nonlinearities with a twofold interplay: on one side, a direct competition between the two nonlinear index wells, i.e., between the two all-optical wave-guiding effects quantified by Eqs.~(\ref{eq:phi2_reor}-\ref{eq:phi2_thermal}) and plotted in Fig.~\ref{fig:comparison_NL_ext} for the extraordinary polarization; on the other side, an increase in temperature  changes the material properties, including the elastic response (via the effective Frank's constant $K$) and the dielectric response according to Eqs.~(\ref{eq:nnor}-\ref{eq:npar}) (see Section \ref{sec:elastic_properties} and Fig.~\ref{fig:K_vs_temperature} in the Appendix for more details) \footnote{In general, optical reorientation also modulates the thermal properties. We neglect this correction consistently with the background orientation $\theta_0=\pi/4$, as $\psi$ remains small with respect to $\theta_0$.}. \\
The IR beam is able to excite a bright soliton with either input polarizations. When the polarization is extraordinary ($y$-polarized beam) reorientation always overcomes thermal effects [see Fig.~\ref{fig:comparison_NL_ext}(b)]. An upper bound for soliton power exists due to the isotropic-to-nematic transition [see Eq.~\eqref{eq:PNI}]. The soliton width does not decrease monotonically with power, owing to a decrease in the reorientational respone via heating  [see Fig.~\ref{fig:comparison_NL_ext}(b)]. The ordinary wave can also excite a soliton of purely thermal origin, even if such solitons are much wider than reorientational ones (tens of microns with respect to a few microns for typical powers, see Fig.~\ref{fig:soliton_w_vs_P}). With reference to the reorientational soliton, the thermal effect increases as the transition to the isotropic phase is approached (Fig.~\ref{fig:index_vs_temperature}). Noteworthy, to ensure the observability of a scalar thermal soliton, the power must remain below the Fr\'eedericksz threshold, the latter given by Eq.~\eqref{eq:Pth} once the waist of the corresponding soliton is used instead of the generic beam width $w$. In Fig.~\ref{fig:soliton_w_vs_P} (center panel) the existence branches where the Fr\'eedericksz threshold is surpassed are marked by dashed lines. \\
Due to the much larger absorption and the different dispersion, self-trapping of green waves strongly differs from the infrared case. First, the extraordinary component never forms a bright soliton owing to the dominant defocusing of thermal origin (Fig.~\ref{fig:comparison_NL_ext}). Conversely, an ordinary-wave thermal soliton can fe formed. Due to the magnitude of the absorption, the soliton can be very narrow (see Fig.~\ref{fig:soliton_NL_alone}), but self-trapping  can take place at very small powers (less than 1mW) and in a very narrow range dependent on the initial temperature $T_0$. Similarly to the IR case, the power upper bound corresponds to the nematic-to-isotropic phase transition  at $T>T_\mathrm{NI}$. The lower bound is associated with the peculiar dispersion of $n_\bot$ at this wavelength: for $T<300.15$~K the ordinary index decreases with temperature (see Fig.~\ref{fig:index_vs_temperature} in the appendix), thus a bright soliton can exist only when the power inverts the sign of the nonlinearity and yields self-focusing. In agreement with this, the existence curves for solitons shift towards lower powers as the initial temperature  increases.          

\section{Evolution of the beam width}

In this section, using Eqs.~\eqref{eq:dynamic_breathing} and (\ref{eq:phi2_reor}-\ref{eq:phi2_thermal}), we use the semi-analytical model of Ref.~\cite{Karimi:2016} to study theoretically how a fundamental Gaussian beam propagates in doped NLC, considering both polarizations separately and two wavelengths, one off-resonance (IR) and one (green) strongly absorbed. As input, we will take a Gaussian beam of input width $w_0$ and possessing a flat phase profile in the input section $z=0$.

\subsection{IR beam alone}

\begin{figure}
\includegraphics[width=0.49\textwidth]{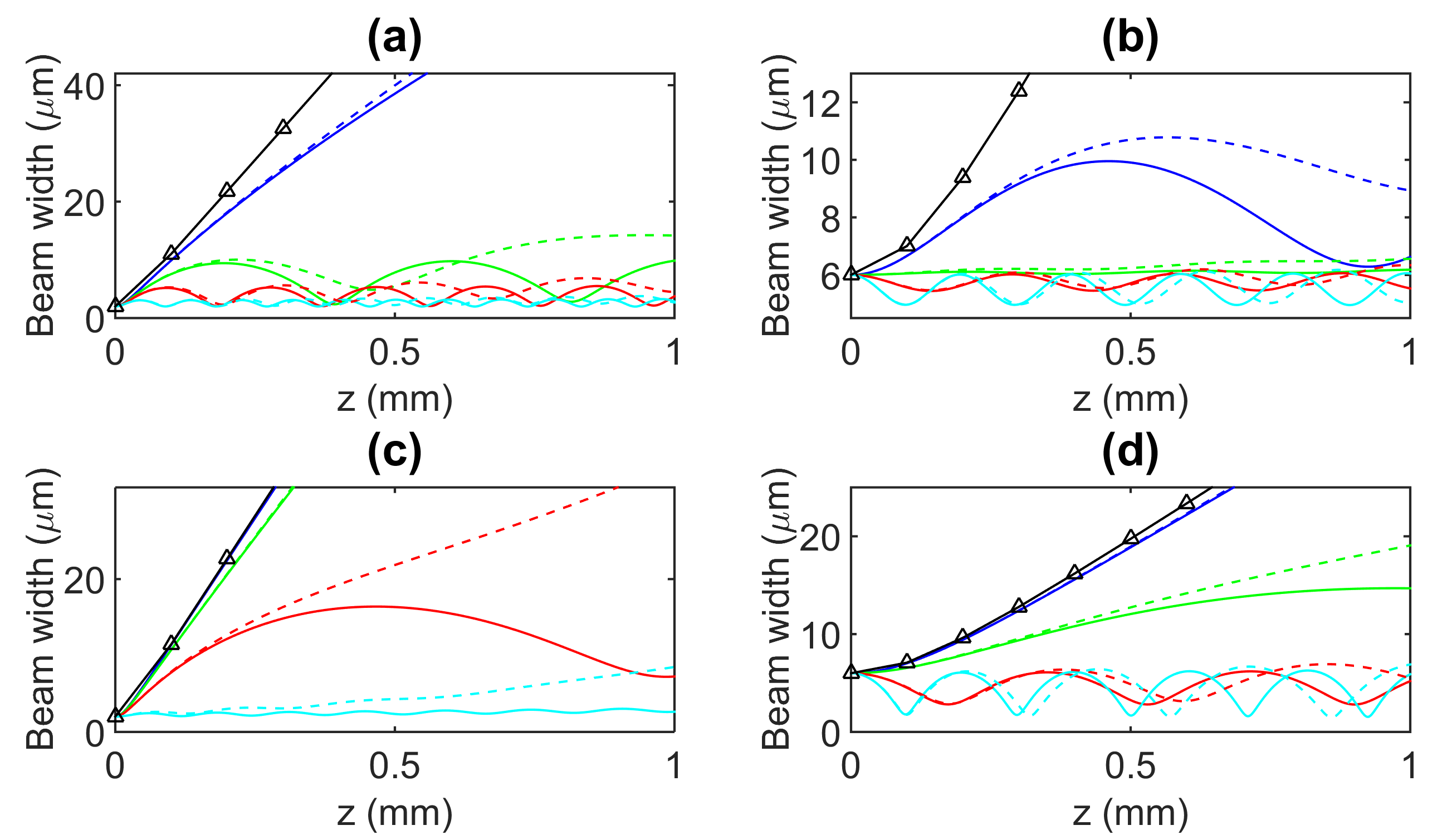}
\caption{Beam width versus $z$ when $\lambda=1064$~nm for $\alpha=10^2$~m$^{-1}$ and an input width $w_0=2~\mu$m (a, c) and $w_0=6~\mu$m (b, d), respectively. (a-b) Extraordinary wave subject to a purely reorientational nonlinearity for input power $P_0=1~$mW (blue lines), 3~mW (green lines), 5~mW (red lines) and 10~mW (cyan lines). (c-d) Ordinary wave subject to a purely thermo-optic nonlinearity for input power $P_0=1$~mW (blue lines), 8~mW (green lines), 30~mW (red lines) and 50~mW (cyan lines). Solid and dashed lines correspond to zero Rayleigh scattering  $\alpha_\mathrm{el}=0$ and to $\alpha_\mathrm{el}=4\times 10^{2}$~m$^{-1}$, respectively. %In every panel solid and dashed lines correspond to $\alpha_\mathrm{el}=0$ and $10^3$~m$^{-1}$, respectively, 
The black line with triangles plots the linear diffraction; the initial temperature $T_0$ is 295.15~K. }
\label{fig:IR_width_vs_z}
\end{figure}

The behavior of the IR beam width versus $z$ was computed numerically by solving Eq.~\eqref{eq:dynamic_breathing} for different input powers and is plotted in Fig.~\ref{fig:IR_width_vs_z}. %The behavior of the extraordinary component is shown in Fig.~\ref{fig:IR_width_vs_z}(a-b). 
For the extraordinary component [Fig.~\ref{fig:IR_width_vs_z}(a-b)], we only accounted for the reorientational index well given by Eq.~\eqref{eq:phi2_reor}, as it is predominant with respect to the thermal effect (see Fig.~\ref{fig:comparison_NL_ext}); in the calculation, however, we included the thermal modulation of the NLC parameters (refractive indices and elastic constants) through absorption. In agreement with previous literature, strong self-focusing yields spatial solitons at a few mW powers. The dynamics of soliton formation depends on the input beam width: self-focusing of wider beams is eased owing to less diffraction \cite{Karimi:2016}. Losses due to Rayleigh scattering (not contributing to thermal heating) affect self-trapping and increase both the average beam width in propagation and the oscillation period (see solid and dashed lines in Fig.~\ref{fig:IR_width_vs_z}). \\
In the ordinary polarization, the beam follows a similar dynamics induced by the thermal nonlinearity, which is the only one active for powers below the Fr\'eedericksz transition (expressed by Eq.~\eqref{eq:Pth}). Even though the power required for self-focusing is much higher because of a lower nonlinearity, the influence of scattering losses and input beam width is analogous to the extraordinary case. %Noteworthy, such model holds valid only for powers below the OFT, the latter .

\subsection{Green beam alone}

\begin{figure}
\includegraphics[width=0.49\textwidth]{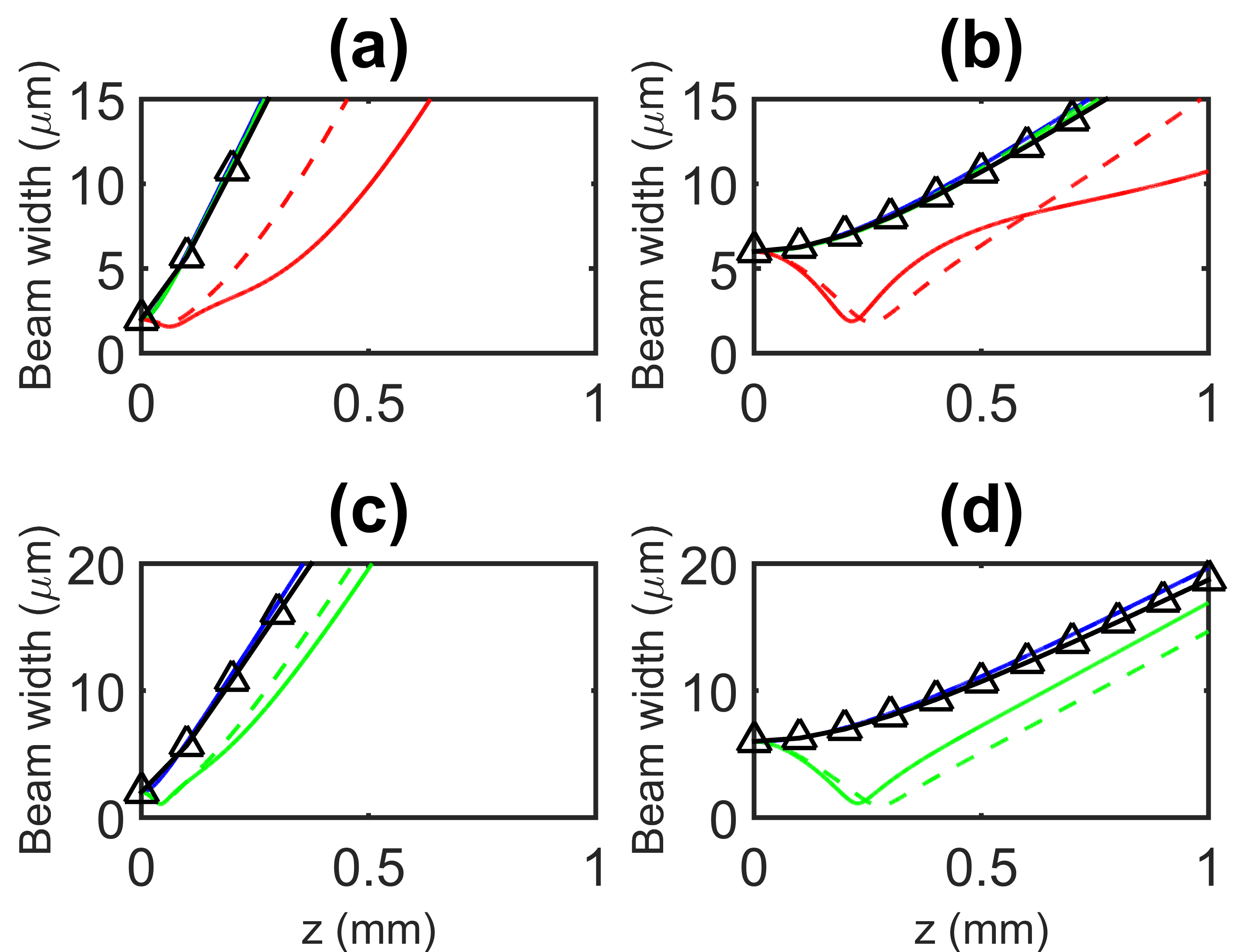}
\caption{Beam width versus $z$ when the input wave is purely ordinary at 532~nm. The thermal absorption $\alpha$ is 9 $\times\ 10^{2}$~m$^{-1}$  (a-b) and  3 $\times\ 10^{3}$~m$^{-1}$ (c-d),  the input beam width $w_0$ is 2$~\mu$m (a, c) and 6$~\mu$m (b, d), respectively. The input power is 0.1~mW (blue lines, a-d), 1.5~mW (green lines, a-d), and 5~mW (red lines, a-b), respectively. In all panels the black line with triangles graphs the linear diffraction; the initial temperature $T_0$ is 298.15~K.}
\label{fig:green_ord_theory}
\end{figure}

\begin{figure}
\includegraphics[width=0.49\textwidth]{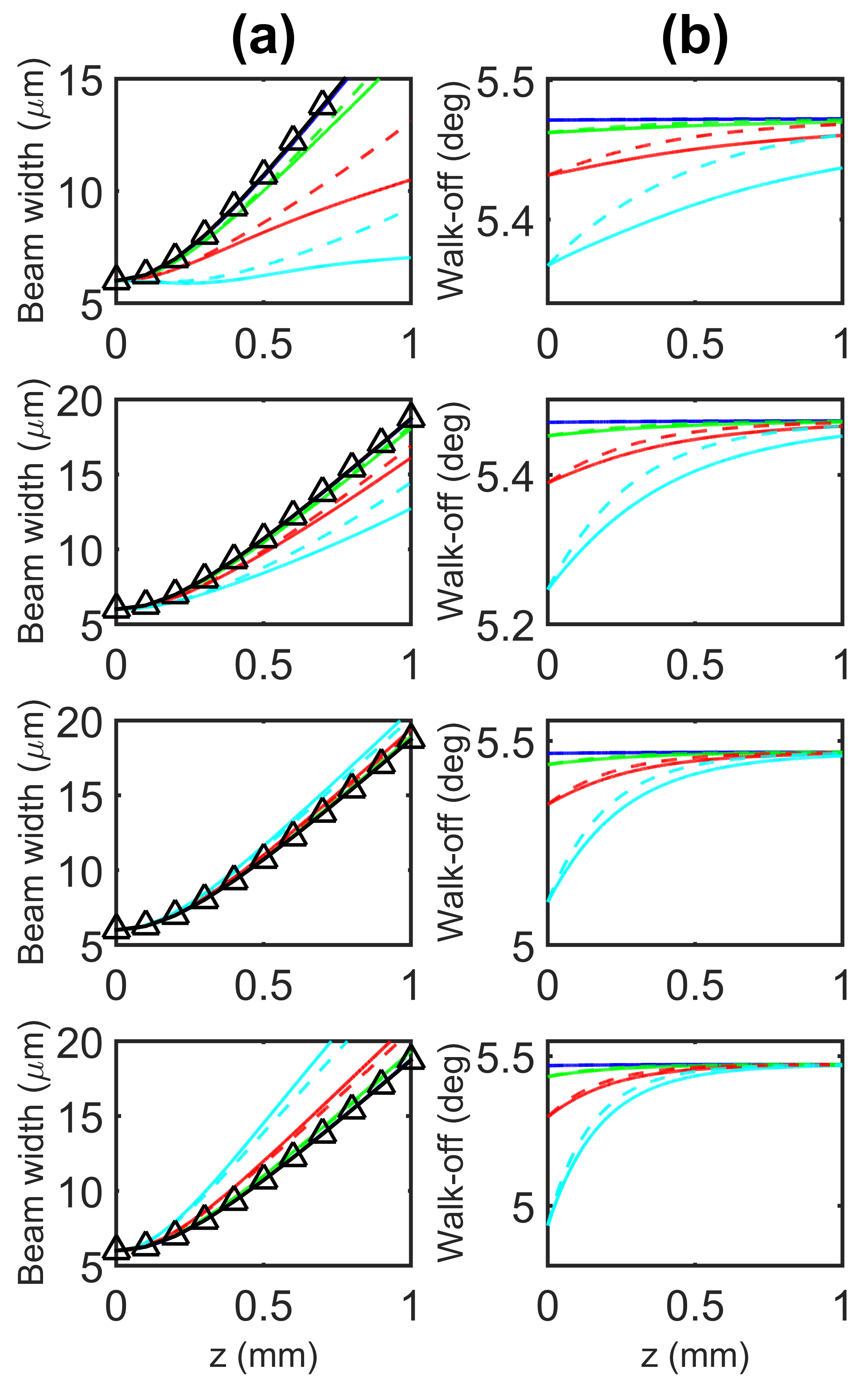}
\caption{Beam width (column a) and walk-off angle (column b) versus $z$ when the input polarization is extraordinary at 532~nm. Thermal absorption $\alpha$ is $1\ \times 10^3$ ~m$^{-1}$,$\ 2\times 10^3$~m$^{-1}$,  $3\times 10^3$~m$^{-1}$ and $4\times 10^3$~m$^{-1}$ from top to bottom, respectively. Input powers are 0.01~mW (blue lines), 0.1~mW (green lines), 0.4~mW (red lines), and 1~mW (cyan lines). Solid and dashed lines correspond	to $\alpha_\mathrm{el}=0$~m$^{-1}$ and $10^3$~m$^{-1}$, respectively. In all panels the black line with triangles refers to linear diffraction; $T_0$ is 298.15~K and $w_0$ is $5~\mu$m.}
\label{fig:green_ext_theory}
\end{figure}

The behavior of the green beam, when the input polarization is ordinary, is plotted in Fig.~\ref{fig:green_ord_theory}. At low powers, small increases in beam divergence are observed owing to the defocusing sign of $dn_\bot/dT$ for $T<300.15$~K (see Fig.~\ref{fig:index_vs_temperature}). At higher powers, the thermal nonlinearity becomes self-focusing. Due to the higher absorption $\alpha$ in the green, self-lensing is stronger near the input interface with respect to the infrared case, but self-trapping fades away more rapidly due to larger losses. In fact, as the absorption $\alpha$ increases (e.g., larger concentration of dopants), the minimum beam width gets smaller but, at the same time, the beam starts to freely diverge after shorter propagation distances $z$. This effect is prominent for small input beam widths [see Fig.~\ref{fig:green_ord_theory}(a,c)]. For wider input beams, self-trapping is more prone to occur and survives on larger distances from the input interface as well [Fig.~\ref{fig:green_ord_theory}(b,d)]. \\
The propagation of the extraordinary green beam is more involved than in the infrared case: due to the large absorption, thermal and reorientational nonlinearities are comparable, and both have to be accounted for simultaneously. Figure~\ref{fig:green_ext_theory}(a) plots beam width versus $z$ for an input width of 2~$\mu$m and  four absorption coefficients $\alpha$. For small absorption, reorientation is dominant and an overall focusing  takes place. Between $\alpha=2\ \times 10^3$~m$^{-1}$ and $\alpha=3\ \times 10^3$~m$^{-1}$, the thermal response overcomes the reorientational nonlinearity, yielding a monotonic increase in  beam divergence versus input power as compared to the linear case, i.e., self-defocusing in agreement with Fig.~\ref{fig:comparison_NL_ext}. Due to the large increase in temperature, an extraordinary beam is also able to change its own trajectory: the light-induced temperature increments $T_m-T_0$ yield a power-dependent change in walk-off via Eqs.~(\ref{eq:nnor}-\ref{eq:npar}). In particular, for $\theta_0=45^\circ$ walk-off decreases with power: the wave-fronts remain unperturbed (the average wave vector remains normal to $\hat{z}$) but the Poynting vector changes direction. Fig.~\ref{fig:green_ext_theory}(b) shows the computed walk-off angle $\delta=\delta(\theta_0=\pi/4,T=T_m)$ versus $z$. For $z\gg1/\alpha_\mathrm{ov}$, the walk-off tends to $\delta=\delta(\theta_0=\pi/4,T=T_0)$, i.e., the beam direction with respect to $z$ corresponds to the linear case. Power-driven variations in walk-off mimic the trend of the maximum temperature $T_m$, with an exponential decay along $z$ with slope given by the overall losses $\alpha_\mathrm{ov}$.

\section{Co-propagation of a weak probe and an intense IR beam}

In this section we will investigate experimentally the light propagation when the impinging wavelength is outside the dye absorption band: in our case  $\lambda=1064$~nm. We launched a linear polarization exciting either ordinary ($x$-polarized) or extraordinary ($y$-polarized) waves.
\begin{figure}
\includegraphics[width=0.5\textwidth]{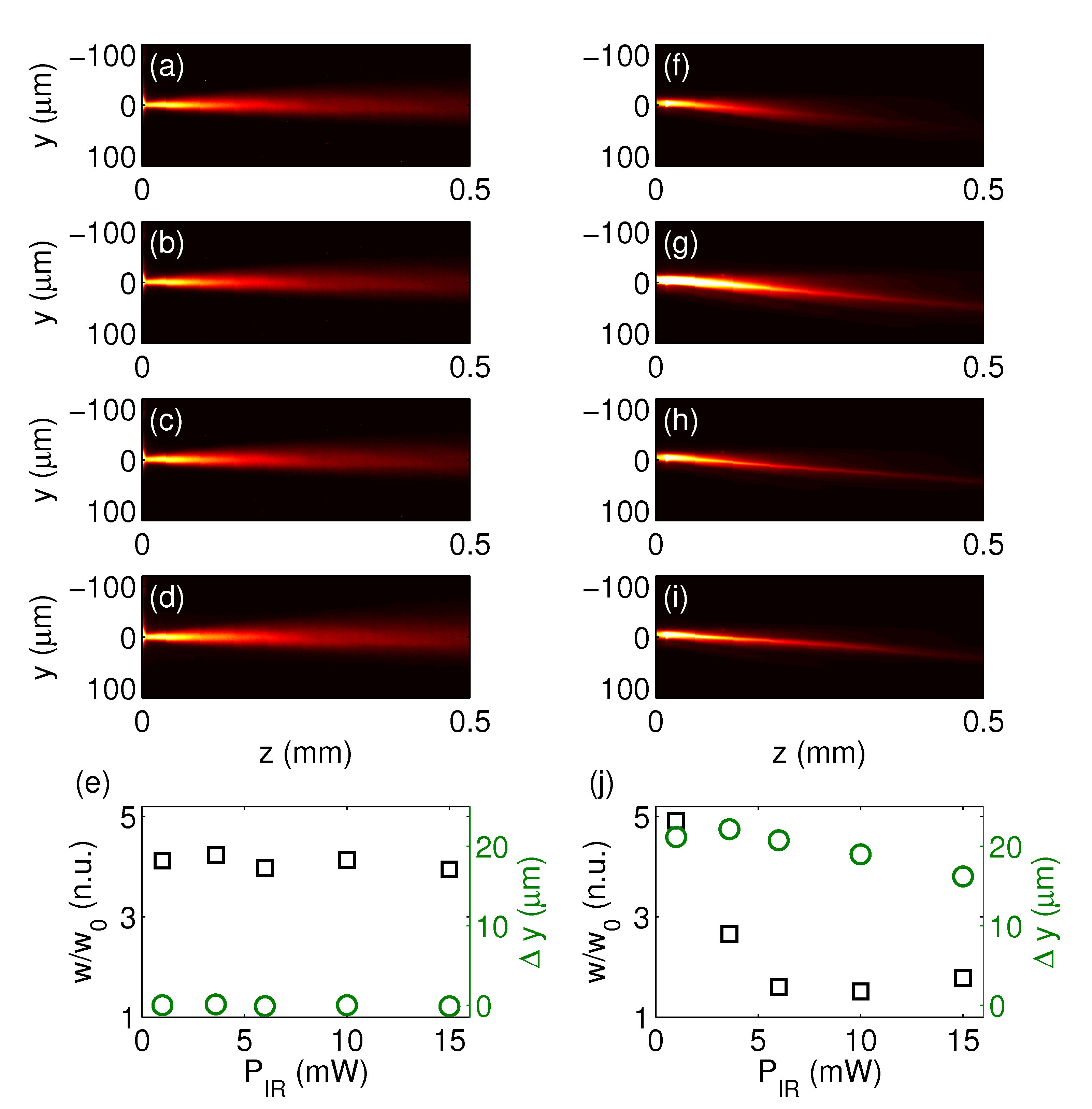}
\caption{Experimentally acquired photographs of a green probe beam co-polarized and co-propagating with an IR beam in the $yz$ observation plane of a planar cell. When the two beams are ordinary waves (a-d), the Fr\'eedericksz transition inhibits reorientation in either width [normalized to the initial value $w_0=w(z=0)$] and Poynting vector; the IR power is $P=1$~mW (a), (b) $3$~mW, (c) $6$~mW, and (d) $10$~mW. Panel (e) summarizes normalized beam width and transverse displacement [$\Delta y \equiv y(z=z_p)-y(z=0)$] due to walk-off in $z=z_p=250~\mu$m. If the beams are extraordinary waves [see panels (f-i)], reorientation causes self-focusing and a nematicon guides the probe when $P=1$~mW (f), $3$~mW (g), $6$~mW (h) and $10$~mW (i). (j) Same as in (e), but for extraordinary waves. The lower limit for the accuracy is about $1~\mu$m and $0.2$ for the position and the normalized width, respectively.}
\label{fig:exp_reorientational}
\end{figure}
Figure~\ref{fig:exp_reorientational} illustrates the propagation of a weak probe (green) beam at $P_{vis}=100~\mu$W, unable to excite either the reorientational nonlinearity or appreciable thermal effects, co-polarized and co-launched with an intense IR beam. The IR beam impinges normally to the sample, both for extraordinary and ordinary polarizations; the probe is launched so that its path in the NLC  overlaps with the IR. Thus, the green beam wave vector in the extraordinary polarization is tilted so that the two waves share the same Poynting vector direction, regardless of dispersion and walk-off (Fig.~\ref{fig:index_vs_temperature}). \\
Let us start with the ordinary polarization. The beam evolution for various IR beam powers is plotted in Fig.~\ref{fig:exp_reorientational}(a-d). Fig.~\ref{fig:exp_reorientational}(e) shows the normalized beam width $w/w_0$ and the beam shift $\Delta y$ with respect to the input section versus the input IR power, measured at a distance $z=z_p=0.25~$mm from the input interface. No appreciable variations are observed, neither in beam size nor in trajectory. Regarding the latter, no changes are expected as the walk-off is zero for this polarization. With reference to the beam size, Fig.~\ref{fig:soliton_w_vs_P} predicts a minimum soliton width of about $7~\mu$m around $P=20~$mW; due to the fact that the input waist is about $3~\mu$m, the ordinary wave is negligibly affected by thermal self-focusing. Such behavior is also confirmed by the calculated width  in Fig.~\ref{fig:IR_width_vs_z}(c) in the presence of scattering losses (dashed lines), where even for $P=50~$mW the beam undergoes spreading at $z_p=0.25$~mm. Furthermore, the Fr\'eedericksz transition, estimated  at powers just below 20~mW (Fig.~\ref{fig:soliton_w_vs_P}), is not observed up to $P_{IR}=50~$mW, when detrimental effects from heating (like formation of isotropic bubbles related to the isotropic-nematic transition in NLC regions) take place. Such discrepancy between theory and experiments can be attributed to inhomogeneities in director distribution at the cell entrance, possibly through the formation of a meniscus at the interface air-NLC. Such imperfections can affect reorientation more than the thermal flow. \\
The evolution of the extraordinary wave is shown in Fig.~\ref{fig:exp_reorientational}(f-i) at four powers. For the sake of a quantitative discussion, Fig.~\ref{fig:exp_reorientational}(j) graphs the normalized beam width $w/w_0$ and lateral shift $\Delta y$ in $z=z_p$. The extraordinary polarized beam at low powers has its Poynting vector at a walk-off angle $\delta(\pi/4)\approx4.5^\circ$ (the theory predicts $4.8^\circ$ at $T=295$~K). Small variations can be observed in the beam trajectories with increasing powers, as lower walk-off is associated to a slightly reduced anisotropy in regions with higher temperatures (see Fig.~\ref{fig:index_vs_temperature}). The width of the probe versus $z$ shows appreciable changes for $P_{IR}>1$~mW due to the waveguide induced by molecular reorientation. 
The beam width at $z=z_p=0.25$~mm is about the same as at the input section when $P_{IR}\approx 4~$mW. This is in good agreement with Fig.~\ref{fig:IR_width_vs_z}(a), where the beam width reacquires the same value inside the NLC cell for powers between 3 and 5~mW. For higher powers, consistently with Fig.~\ref{fig:IR_width_vs_z}(a)  the width measured in $z_p$ oscillates around the size of a shape-preserving soliton, between 1.5~$\mu$m and 4~$\mu$m for 5~mW$~<P<~40~$mW according to the leftmost panel in Fig.~\ref{fig:soliton_w_vs_P}. 

\section{Light propagation at wavelength within the absorption band of the dye}

\begin{figure}
\includegraphics[width=0.5\textwidth]{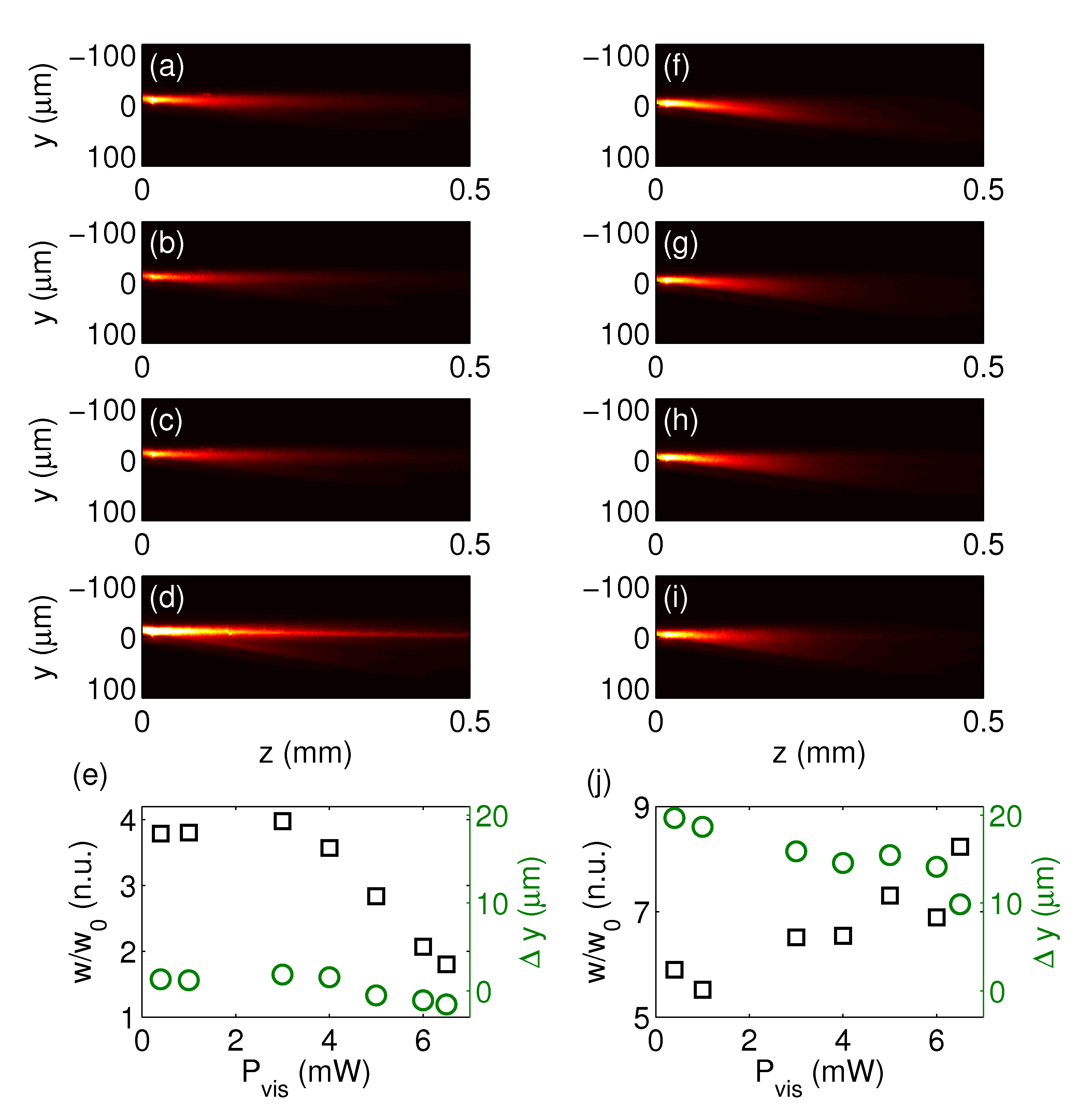}
\caption{Experimentally acquired photographs of the propagating visible beam (no IR). Ordinary wave propagation for $P=0.4$~mW (a), $3$~mW (b), $4$~mW (c), and $6$~mW (d), respectively, showing thermal self-focusing. (e) Normalized beam width $w/w_0$ and beam displacement $\Delta y$ versus input power $P$. Propagation of extraordinary wave beam for $P=0.4$~mW (f), $3$~mW (g), $4$~mW (h), and $6$~mW (i), respectively: the extraordinary undergoes self-defocusing and power-dependent changes in  trajectories. (j) Normalized width $w/w_0$ and transverse displacement $\Delta y$ of the extraordinary beam versus input power. The lower limit for the accuracy is about $1~\mu$m and $0.2$ for the position and the normalized width, respectively.}
\label{fig:exp_thermal}
\end{figure}
When the green probe propagates by itself in the cell, the observed behavior drastically changes, as predicted. Let us first analyze the evolution of the ordinary component, shown in Fig.~\ref{fig:exp_thermal}(a-d) with quantitative features in Fig.~\ref{fig:exp_thermal}(e). When the green power is above 3~mW, the beam experiences thermal self-focusing; for $P=6.5~$mW, the beam shrinks down to a size close to the initial value; further increases in power eventually lead to the nematic-isotropic transition. This agrees well with the theoretical results plotted in Fig.~\ref{fig:green_ord_theory}(a). Finally, the transverse position of the ordinary beam slightly depends on the input power, due to slight variations in wave vector near the NLC-air interface and probably to a meniscus (as speculated in the previous case above), with slightly asymmetric boundary conditions near the beam entrance.  \\ 
Considering the extraordinary polarization, graphed in Fig.~\ref{fig:exp_thermal}(f-j), a quasi-linear increase of  beam width versus input power is observed [see Fig.~\ref{fig:exp_thermal}(j)], associated with defocusing for $P>1~$mW. The lateral displacement $\Delta y$ at $P=6.5~$mW is twice smaller than in the linear regime (low powers). Such a deflection cannot be ascribed solely to thermally-induced changes in walk-off: as visible in Fig.~\ref{fig:green_ext_theory}, in fact, such gradual changes should be limited to about 0.5$^\circ$. These experimental results prove that the wave vector undergoes a nonlinear deflection as well, in analogy with what we previously described for green ordinary waves.

 %and starting from $P=0.4$~mW, an increase of the divergence is observed for the walking-off beam as the power is risen up to $P=6$~mW (limit due to the formation of isotropic domains), as shown in Fig.~\ref{fig:exp_thermal}(j). Correspondingly, as expected, the angular displacements between Poynting and wave vectors reduces, as a consequence of the reduction of the overall medium anisotropy. 

%The same experiment was repeated for $P_{IR}=8mW$, under a stronger IR self-focusing regime. Results, shown in fig.??, confirm that the competing nonlinearities allow a control of both trajectory (position) and width (confinement) of the IR self-confined wave, depending of the relative amount of power. Due to the higher confinement, the trajectory bending is limited to the first $500\mu m$, when the defocusing response is stronger, while the fastest breathing is less affected by the green beam. 

\section{Conclusions}

We investigated the interplay between two competing optical nonlinearities in NLCs, the reorientational and the thermal one.
In the highly nonlocal limit and for small light-induced rotations of the director, the two effects share the same profile for the nonlinear perturbation. As a direct consequence, in the highly nonlocal regime the relative weight of the two responses does not depend on the the spatial profile of the input beam and it is spatially uniform across the NLC layer. We discussed the interplay between the two nonlinearities and its dependence on initial temperature and material absorption. In dye-doped NLCs, the latter can modulate the strength of the opto-thermal effects, usually negligible in undoped NLCs. We showed that there are two different regime for the interplay between the two nonlinearities: i) a direct competition on forming the overall nonlinear index well; ii) a modulation of the parameters determining the NLC response to light. With reference to the latter, we showed that light-induced changes in temperature significantly affect the formation of reorientational solitons even in the undoped case, including the existence of solitons in a limited power range and a non-monotonic soliton width versus power. \\
Our findings widen the perspective on the unique optical properties of NLCs, the latter being an ideal workbench for the study of nonlinear optics and the interaction between nonlinearities \cite{Maucher:2016,Jung:2017}. Future developments include the simultaneous propagation of two beams at different wavelengths and with different profiles \cite{Laudyn:2015}. Generalizations to nonlinearities acting on distinct time scales can be envisaged as well when using pulsed sources  \cite{Burgess:2009}. Our results, together with semi-analytic models accounting for self-lensing and its dynamics in propagation, show how nonlinear optics in extended samples is an important tool for the complete characterization of NLC mixtures. 
  
 %Theoretical and experimental results shows that is possible to contemporaneously excite the two kind of nonlinear behaviors using two different beams, so that the final propagation is resulting from the interplay of the two beams. In particular we shows how the thermal nonlinearity (in this case the lowest one) induced by one beam can be used to affect both linear and nonlinear properties of the NLC sample, in order to control the propagation properties of the second beam, i.e. trajectory and confinement properties. We think this findings open the way to a number of phenomena involving the synergetic effect of different nonlinear mechanisms in NLC, favouring the observation of novel mechanisms for spatial soliton formation and control. 

\section*{Appendix}

\subsection{Linear optical parameters of the mixture 6CHBT}

\begin{figure}
\includegraphics[width=0.47\textwidth]{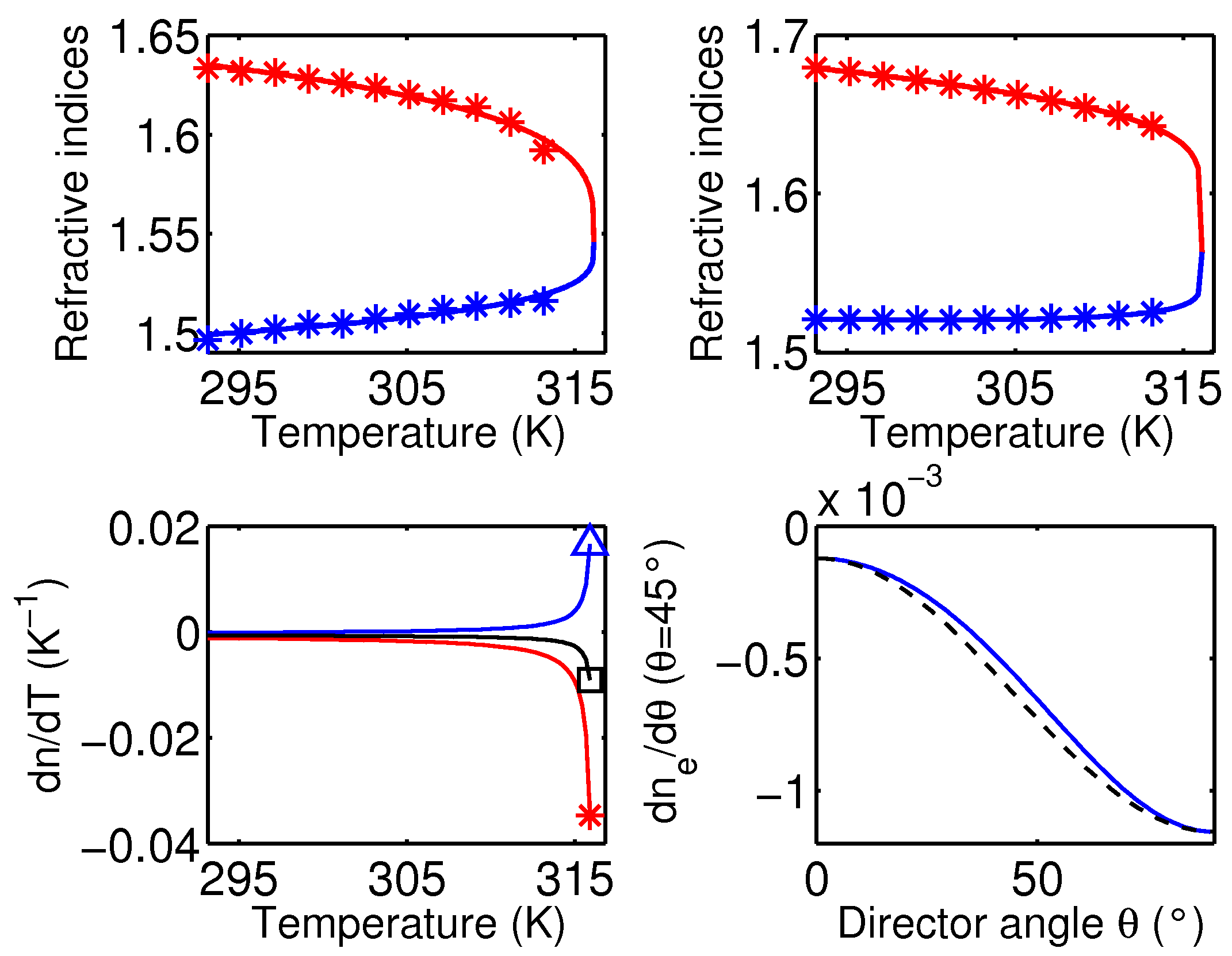}
\caption{ Refractive indices $n_\|$ (red lines, upper curves) and $n_\bot$ (blue lines, bottom curves) versus temperature for (a) $\lambda=1064~$nm and (b) $\lambda=532~$nm; solid lines and symbols are fits and measurements, respectively. (c) Derivatives with respect to temperature of $n_\|$ (red line with star), $n_\bot$ (blue line with triangle) and $n_e(\theta=\pi/4)$ (black line with square) as computed from Eq.~(\ref{eq:sign_thermal}) at $\lambda=532~$nm. (d) Derivative of $n_e$ with respect to the director angle $\theta$ versus $\theta$ for $\lambda=532~$nm and $T=293~$K. Blue solid and black dashed lines are the exact expression (\ref{eq:dne_dT_exact}) and the corresponding approximation (\ref{eq:sign_thermal}), respectively.}
\label{fig:index_vs_temperature}
\end{figure}

Figure \ref{fig:index_vs_temperature} (a-b) shows the comparison between actual measurements and fitting curves given by Eqs.~(\ref{eq:nnor}-\ref{eq:npar}) for the two wavelengths we used in this work. A best-fit procedure provides $A=1.5295$, $B=-5.1364\times 10^{-5}$, $(\Delta n)_0=0.2767$ and $\beta=0.2719$ for $\lambda=1064~$nm, whereas we find $A=1.7098$, $B=4.6545\times 10^{-4}$, $(\Delta n)_0=0.2345$ and $\beta=0.1483$ for $\lambda=532~$nm. Having ascertained the quality of the fitting curves, the latter can be used to compute the derivative of $n_\|$ and $n_\bot$ with respect to temperature, as shown in Fig. \ref{fig:index_vs_temperature}(c). Finally, Fig.~\ref{fig:index_vs_temperature} provides a  direct comparison between the approximated formula~(\ref{eq:sign_thermal}) and the exact expression
\begin{equation}
  \frac{dn_e}{d\theta}=\left(\frac{\cos^2\theta}{n_\bot^2} + \frac{\sin^2\theta}{n_\|^2} \right)^{-\frac{3}{2}} \left[\frac{\cos^2\theta}{n_\bot^3} \frac{dn_\bot}{dT} + \frac{\sin^2\theta}{n_\|^3} \frac{dn_\|}{dT} \right]. \label{eq:dne_dT_exact}
\end{equation}

\subsection{Elastic properties of the mixture 6CHBT}
\label{sec:elastic_properties}
When anisotropy of the elastic properties of the NLC is accounted for, the reorientation induced by an optical wavepacket invariant along $z$ is given by

\begin{align}
  &(K_1\cos^2\theta + K_3 \sin^2\theta) \frac{\partial^2 \theta}{\partial y^2} + K_2 \frac{\partial^2 \theta}{\partial x^2} + \nonumber\\ &(K_3-K_1)\sin(2\theta) \left(\frac{\partial \theta}{\partial y} \right)^2 + \frac{\epsilon_0\epsilon_a}{4} \sin\left[2(\theta-\delta)\right]|E|^2=0. 
\end{align}
\begin{figure}
\includegraphics[width=0.47\textwidth]{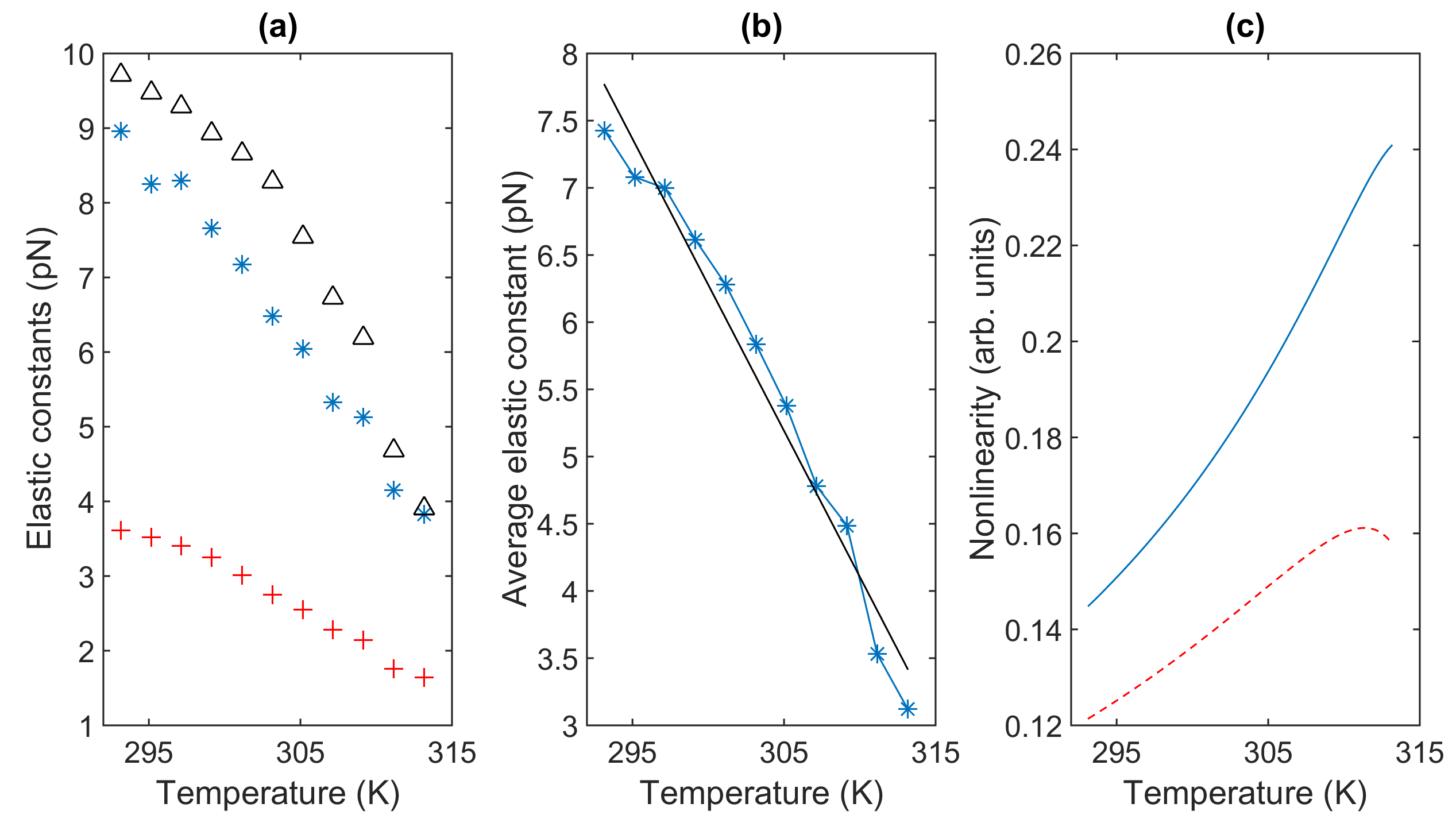}
\caption{ (a) Measured elastic constants $K_1$ (blue stars), $K_2$ (red crosses) and $K_3$ (black triangles) versus temperature. (b) Average elastic constant $K_\mathrm{av}$ versus temperature (blue symbols) computed from the data shown in panel (a); the black solid line is the corresponding linear interpolation. (c) Behavior of the reorientational nonlinearity $\epsilon_0\epsilon_a/(4K_\mathrm{av})$ versus temperature for $\lambda=1064~$nm (red dashed line) and 532nm (blue solid line). }
\label{fig:K_vs_temperature}
\end{figure}
For small optical reorientations the term proportional to $(\partial \theta/\partial y)^2$ can be neglected. Thus, for an initial angle $\theta_0=\pi/4$ an effective elastic constant equal to the average of the three elastic constants can be assumed. The elastic constants for 6CHBT are plotted in Fig.~\ref{fig:K_vs_temperature}(a), as measured by all-optical methods \cite{Klus:2014}. All three elastic constants decrease monotonically versus temperature. Fig.~\ref{fig:K_vs_temperature}(b) shows  the average Frank's constant $K_\mathrm{av}=\left(K_1+K_2+K_3\right)/3$: for temperatures not too close to the transition $T_\mathrm{NI}$, $K_\mathrm{av}$ can be satisfactorily approximated by a linear polynomial with coefficients $K_\mathrm{av}=-0.2173~T+71.4828$. Finally, Fig.~\ref{fig:K_vs_temperature}(c) graphs the ratio of the optical anisotropy and the average elastic constant. All-optical reorientation is stronger at shorter wavelengths and increases with temperature up to $T=40^\circ$C.  

\subsection{Width of solitons supported by a single nonlinearity}

In the highly nonlocal approximation, fundamental solitons feature a Gaussian profile with waist determined by $[2/(n_0 k_0^2 |\phi_2|)]^{1/4}$, where $n_0$ is the unperturbed refractive index and $k_0$ the vacuum wave number. From Eq.~(\ref{eq:phi2_reor}), the existence curve of a soliton in the perturbative regime and due to a reorientational nonlinearity alone is 
\begin{equation}   \label{eq:wsol_reor}
  w^\mathrm{sol}_\theta(P)=\frac{1}{k_0} \sqrt{\frac{8 \eta\pi  K \cos^2\delta_0}{ Z_0\epsilon_0 \epsilon_a \sin[2\left(\theta_0-\delta_0 \right)] \left.\frac{dn_e}{d\theta}\right|_{\theta_0}}} \frac{1}{\sqrt{P}},
\end{equation}
whereas the thermal self-focusing provides an ordinary-wave soliton  with
\begin{equation}   \label{eq:wsol_thermal}
  w^\mathrm{sol}_T(P)= \frac{1}{k_0} \sqrt{\frac{4\eta\pi\kappa}{n_\bot \alpha_\text{ord} \frac{dn_\bot}{dT}}} \frac{1}{\sqrt{P}}.
\end{equation}
\begin{figure}
\includegraphics[width=0.47\textwidth]{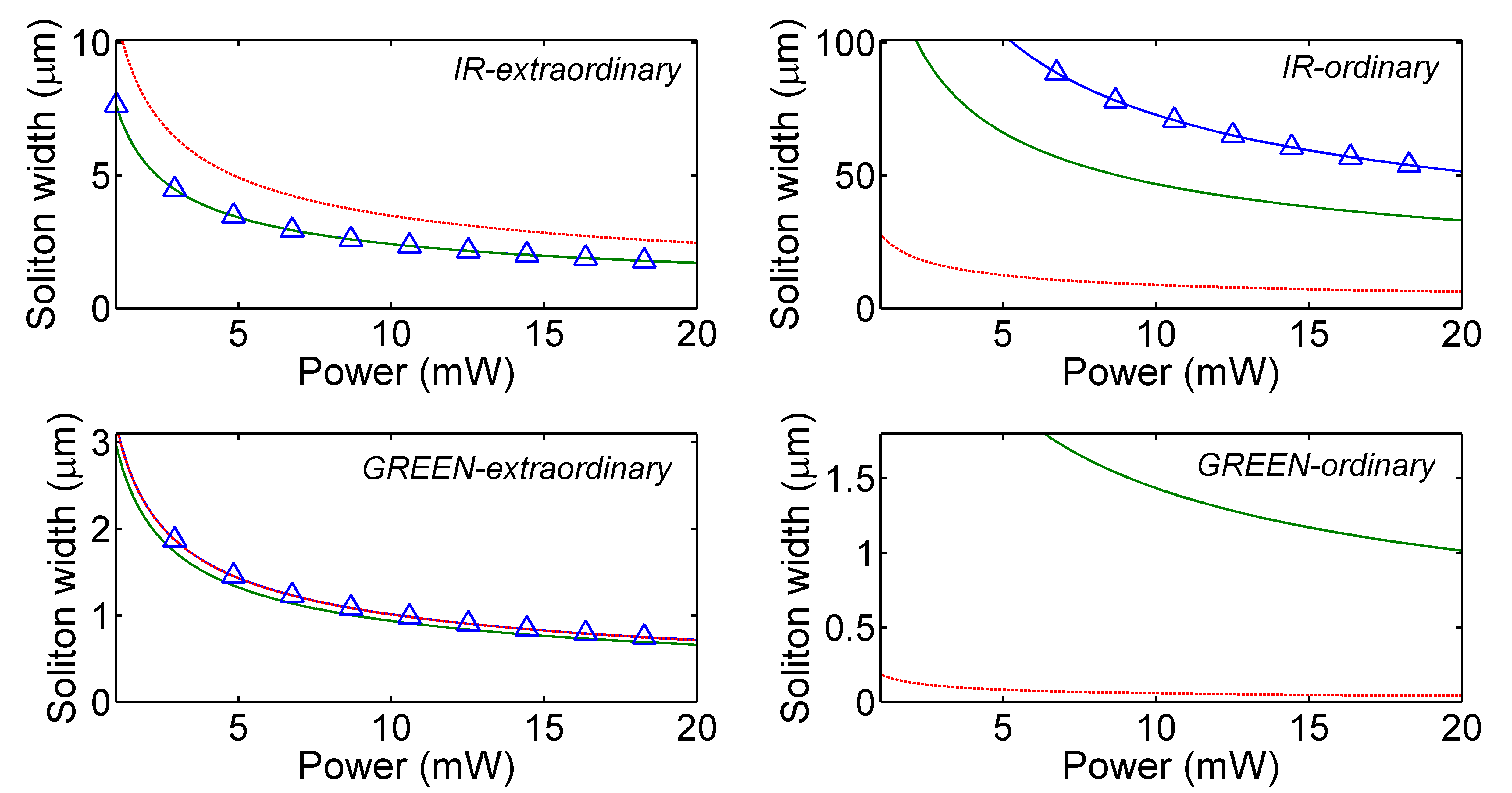}
\caption{ Soliton existence curves in the plane width-power for  extraordinary  (left, reorientational nonlinearity) and  ordinary (right, thermal nonlinearity) waves, as predicted by Eqs.~(\ref{eq:wsol_reor}-\ref{eq:wsol_thermal}). The first and second rows correspond to 6CHBT doped with Sudan Blue for $\lambda=1064$~nm and 532~nm, respectively, with $\theta_0=\pi/4$. Blue lines with triangles, green solid lines and dashed red lines correspond to sample temperatures of $T_m=295~$K, 305~K and 315~K, respectively. The ordinary-wave green soliton at $T=295~$K does not exist because at low temperatures as the thermal response is defocusing, according to Fig.~\ref{fig:index_vs_temperature}.}
\label{fig:soliton_NL_alone}
\end{figure}
The soliton width versus power as predicted by Eqs.~(\ref{eq:wsol_reor}-\ref{eq:wsol_thermal}) is plotted in Fig.~\ref{fig:soliton_NL_alone} for a fixed sample temperature. The interaction between the two nonlinearities is neglected, considering only the reorientational effect for the extraordinary wave, and the thermal effect for the ordinary wave. For the infrared beam, corresponding to $\alpha\approx 10^{2}~$m$^{-1}$, the reorientational soliton is much narrower than the thermal one; the extraordinary self-trapped wave widens as the temperature approaches the transition value $T_\mathrm{NI}$. The size of the ordinary-wave thermal soliton is comparable with the extraordinary-wave soliton (i.e., a few microns) when the sample is close to the nematic-isotropic transition. For the green component, owing to the large absorption, the roles are inverted. Green ordinary solitons can theoretically reach sub-wavelength size owing to the large absorption, and this effect becomes more marked as the temperature approaches the nematic-to-isotropic transition $T_\mathrm{NI}$. In actual samples these waves are ruled out by the large losses associated with $\alpha$ and by the fact that strong absorption destroys the nematic phase. 

\section*{Acknowledgements}

A.A. and G.A. thank the Academy of Finland for support through the Finland Distinguished Professor grant No. 282858. Funding from the National Science Centre of Poland under grant agreement DEC-2012/06/M/ST2/00479 is gratefully  acknowledged. 

%\section*{Author contributions statement}

%G.A. with M.Ka. and U.L. conceived and organized the experiments;  U.L. and M.Kw. conducted the experiments with the help of A.P.; A.A. and A.P. analysed the results and prepared theory and models; A.P., A.A. and G.A. prepared the manuscript.  All authors reviewed the manuscript. 

%\bibliography{references}

%merlin.mbs apsrev4-1.bst 2010-07-25 4.21a (PWD, AO, DPC) hacked
%Control: key (0)
%Control: author (0) dotless jnrlst
%Control: editor formatted (1) identically to author
%Control: production of article title (0) allowed
%Control: page (1) range
%Control: year (0) verbatim
%Control: production of eprint (0) enabled
%

%Manual citation list
%\begin{thebibliography}{1}
%\bibitem{Zhang:14}
%Y.~Zhang, S.~Qiao, L.~Sun, Q.~W. Shi, W.~Huang, %L.~Li, and Z.~Yang,
 % \enquote{Photoinduced active terahertz metamaterials with nanostructured
  %vanadium dioxide film deposited by sol-gel method,} Opt. Express \textbf{22},
  %11070--11078 (2014).
%\end{thebibliography}

\end{document}